\documentclass[onecolumn,english,aps,floatfix,superscriptaddress,amsfonts,amssymb,nofootinbib, pra]{revtex4}
\pdfoutput=1
\usepackage{float}
\usepackage{fancyhdr}
\usepackage[dvipsnames]{xcolor}
\usepackage{graphicx}
\usepackage{epsfig} 
\usepackage{amssymb}
\usepackage{mathrsfs}
\usepackage{amsmath,amsfonts}
\usepackage{booktabs}
\usepackage{hyperref}
\usepackage{pgfplotstable}
\usepackage[footnotesize,singlelinecheck=off,justification=raggedright]{caption}
\usepackage{multirow}
\usepackage{youngtab}
\usepackage{tikz}
\usetikzlibrary{shapes}

\def\bra#1{\mathinner{\langle{#1}|}}
\def\ket#1{\mathinner{|{#1}\rangle}}

\def\Bra#1{\left<1>}
\DeclareMathAlphabet{\mathbbmsl}{U}{bbm}{m}{sl}
{\catcode`\|=\active\gdef\Braket#1{\left<\mathcode`\|"8000\let|\bravert {#1}\right>}}
\def\bravert{\egroup\,\vrule\,\bgroup}
\def\sign{\mathop{\mathrm{sign}}\nolimits}
\def\arg{\mathop{\mathrm{arg}}\nolimits}

\def\ii{{\rm i}}
\def\ee{{\rm e}}

\def\e{{\mathrm e}}

\DeclareRobustCommand\mydiamond{\raisebox{0pt}{\tikz{\node[draw,scale=0.65,diamond,fill=white](){};}}}
\DeclareRobustCommand\mycircle{\raisebox{0pt}{\tikz{\node[draw,scale=0.65,circle,fill=white](){};}}}
\DeclareRobustCommand\mycirclefilled{\raisebox{0pt}{\tikz{\node[draw,scale=0.65,circle, fill=myred](){};}}}
\DeclareRobustCommand\mysolidline{\raisebox{0pt}{\tikz[baseline]{\draw[very thick] (0,.5ex)--++(0.75,0);}}}
\DeclareRobustCommand\mydashedlinea{\raisebox{0pt}{\tikz[baseline]{\draw[very thick, dashdotted] (0,.5ex)--++(0.75,0);}}}
\DeclareRobustCommand\mydashedlineb{\raisebox{0pt}{\tikz[baseline]{\draw[very thick,  dotted] (0,.5ex)--++(0.75,0);}}}
\DeclareRobustCommand\mydashedlinec{\raisebox{0pt}{\tikz[baseline]{\draw[very thick, dashdotdotted] (0,.5ex)--++(0.75,0);}}}
\DeclareRobustCommand\mydashedlined{\raisebox{0pt}{\tikz[baseline]{\draw[very thick, dashed] (0,.5ex)--++(0.75,0);}}}

\definecolor{myred}{HTML}{d7191c}
\definecolor{mygreen}{HTML}{4daf4a}
\definecolor{myviolet}{HTML}{984ea3}
\definecolor{myorange}{HTML}{ff7f00}
\definecolor{myred2}{HTML}{e41a1c}
\definecolor{myblue}{HTML}{377eb8}

\begin{document}

\title{Emptiness formation probability and Painlev\'e V equation in the XY
  spin chain}
\author{Filiberto Ares }
\affiliation{International Institute of Physics, UFRN,
Campos Universit\'ario, Lagoa Nova 59078-970 Natal, Brazil}
\author{Jacopo Viti}
\affiliation{International Institute of Physics, UFRN,
Campos Universit\'ario, Lagoa Nova 59078-970 Natal, Brazil}
\affiliation{ECT, UFRN,
Campos Universit\'ario, Lagoa Nova 59078-970 Natal, Brazil}
\date{\today{}}

\begin{abstract}
We reconsider the problem of finding $L$ consecutive down spins in the ground state of the XY chain, a quantity known as the Emptiness Formation Probability. Motivated by new developments in the asymptotics of Toeplitz determinants, we show how the crossover between the critical and off-critical behaviour of the Emptiness Formation Probability is exactly described by a $\tau$ function of a Painlev\'e V equation. Following a recent proposal, we also provide a power series expansion for the $\tau$ function  in terms of irregular conformal blocks of a Conformal Field Theory with central charge $c=1$. Our results are tested against lattice numerical calculations, showing excellent agreement. We finally rediscuss the free fermion case where the Emptiness Formation Probability is characterized by a Gaussian decay for large $L$.

\end{abstract}
\maketitle

\maketitle

\section{Introduction}\label{sec:intro}
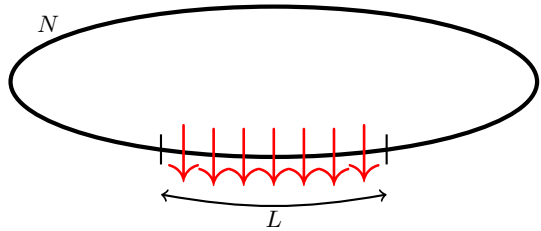
\begin{figure}[t]
 \begin{center} 
 \begin{tikzpicture}
  \draw[ultra thick] (0,0) ellipse (3.5cm and 1cm);
  \node[red] at (0,-1) {\Huge{$\downarrow$}};
  \node[red] at (0.4,-1) {\Huge{$\downarrow$}};
  \node[red] at (-0.4,-1) {\Huge{$\downarrow$}};
  \node[red] at (-0.8,-1) {\Huge{$\downarrow$}};
  \node[red] at (0.8,-1) {\Huge{$\downarrow$}};
  \node[red] at (1.2,-0.95) {\Huge{$\downarrow$}};
  \node[red] at (-1.2,-0.95) {\Huge{$\downarrow$}};
  \draw[thick] (-1.5,-1.1)--(-1.5,-0.7);
  \draw[thick] (1.5,-1.1)--(1.5,-0.7);
  \draw[<->, thick] (-1.5, -1.5) to[out=-10,in=190] (1.5,-1.5);
  \node[below] at (0,-1.6) {$L$};
  \node[below] at (-3,1) {$N$};
 \end{tikzpicture}
\end{center} 
 \caption{Schematic representation of a configuration of $L$ consecutive down spins in a state of a quantum spin chain with total length $N$. The EFP is the probability that such a configuration is realized in the ground state.} 
 \label{typical} 
 \end{figure}

The Emptiness Formation Probability (EFP) represents perhaps the simplest correlator that can be calculated in a quantum spin chain~\cite{Korepin-book}. It is defined as the probability to find a string of $L$ consecutive down spins in the ground state of the system, whose total length is $N$. For a pictorial representation of the problem setting,  see fig.~\ref{typical} in which, conventionally, the direction of the spin is chosen along the $z$-axis. Moreover, from now on, we will focus only on the thermodynamic limit $N\rightarrow\infty$.

The EFP was introduced in~\cite{Korepin1, Essler}, in their study of correlation functions in the spin-$1/2$ XXZ chain. In particular, in~\cite{Essler} it was proposed a Fredholm determinant representation in the thermodynamic limit, through Bethe Ansatz methods~\cite{Korepin-book}. Subsequent developments, including exact asymptotics for large $L$ and finite-$N$ effects, were analyzed in the references~\cite{Shiroishi, Abanov-Korepin, Kitanine1, Kitanine2, Lukyanov, Kozlowsky, Cantini}.
 The large-$L$ asymptotics of the EFP is particularly interesting in the critical XXZ chain and for zero external field, where it is characterized by a Gaussian decay with $L$~\cite{Shiroishi, Abanov-Korepin, Lukyanov}. The latter can be ascribed to the presence of a $U(1)$ symmetry, preserving the total magnetization along the $z$-axis. In a qualitative bosonized description, the behaviour of the EFP is dominated by action configurations~\cite{Abanov-Korepin, Abanov-lectures} that contain a frozen area of order $L^2$ surrounding the string. The Gaussian behaviour of the EFP in the XXZ spin chain has been also related~\cite{Stephan, Allegra} to the phenomenon of phase separation and the existence of arctic curves in the six-vertex model with domain wall boundary conditions~\cite{Colomo1, Colomo2}.

In systems for which the total magnetization is not conserved, the EFP is expected to decay exponentially fast for large $L$. In such a case indeed~\cite{Stephan, M1, M2, M3}, the string of down spins should renormalize to a conformal invariant boundary condition~\cite{Cardy}, whose contribution to the free energy is at most linear with $L$. As we will review in the first part of the paper, this expectation has been indeed supported analytically in the XY spin chain in~\cite{Abanov, Franchini}. Since the XY spin chain will be also the subject of this study, it is convenient to state  already here  its Hamiltonian~\cite{LSM}

\begin{equation}\label{xy_chain}
  H=\sum_{n=1}^N\left(\frac{1+\gamma}{2}\sigma_n^x\sigma_{n+1}^x+
  \frac{1-\gamma}{2}\sigma_n^y\sigma_{n+1}^y\right)-h\sum_{n=1}^N\sigma_n^z,
\end{equation}
being $\sigma_{n}^{\mu}$, $\mu=x,y,z$ Pauli matrices, the parameter $\gamma$ is called anisotropy and $h$ is the transverse magnetic field. The model is  technically easier~\cite{LSM} to solve than the XXZ spin chain and it reduces to a particular case of it when $\gamma=0$. It has moreover the advantage of being a prototype for a quantum phase transition, which is triggered by the transverse field. In particular, in the XY chain, the EFP can be written as a determinant of  a Toeplitz matrix. Beside the expected exponential decay, in~\cite{Franchini}  it was  observed that the EFP, $\mathcal{P}(L, h,\gamma)$, contains power-law prefactors along the critical lines $h=\pm 1$, $\gamma\not=0$. In brief, exploiting known theorems for the asymptotics of Toeplitz determinants~\cite{Szego, Fisher, Widom, Basor1, Basor2, Basor3}, ref.~\cite{Franchini} obtained that
\begin{equation}
\label{FH_intro}
\mathcal{P}(L, h=\pm 1,\gamma\not=0)\simeq L^{-\nu} e^{-|A|L}
\end{equation}
at quantum critical point, while away from it the EFP decays as
\begin{equation}
\label{Szego_intro}
\mathcal{P}(L, h\not=1,\gamma\not=0)\simeq e^{-|A| L}.
\end{equation} 
In~\cite{Stephan}, it was also later clarified how the exponent $\nu$ in \eqref{FH_intro} could be related to the central charge of the underlying Conformal Field Theory (CFT) describing the quantum critical point. For the XY chain with
non-zero anisotropy, such a theory is the critical Ising field theory with central charge $c=1/2$~\cite{BPZ} and $\nu=1/16$.

In this paper we come back on the problem of analyzing the EFP in the XY chain and study analytically the crossover between the critical asymptotics given in~\eqref{FH_intro} and the off-critical  in~\eqref{Szego_intro} in the limit of large $L$. Thus we provide a full characterization of the EFP on the whole phase space of the XY spin chain, complementing the results in~\cite{Abanov, Franchini}.

From the technical point of view our analysis is based on the application of a  recent mathematical theorem by~\cite{Claeys}, in the  theory of Toeplitz determinants. As will be discussed in detail, the symbol associated to the Toeplitz matrix of the EFP  has an emergent Fisher-Hartwig singularity in the limit $h\rightarrow \pm 1$. In particular, specializing the results in~\cite{Claeys}, the EFP  is directly related to the so-called $\tau$ function~\cite{Jimbo} of a Painlev\'e V equation. The mechanism of the emergence of Painlev\'e transcendents in such a context is analogous to the order/disorder transition in the 2d Ising model~\cite{McCoy-book}. In particular the theorem proven in~\cite{Claeys} also applies to the, rather well known example~\cite{McCoy, Tracy-Ising}, of the two-point function of the order parameter in the 2d Ising model. Painlev\'e transcendents are  recurrent in statistical physics and especially for free fermionic models, see for instance~\cite{Tracy-review}.

The rest of the paper is then organized as follows: 
In section~\ref{sec:efp} we will review, from a slightly different perspective the results in~\cite{Abanov, Franchini}; in section~\ref{sec:transition} we will describe the possible crossovers between the different regimes in the phase space of the XY chain; in section~\ref{sec:interpolation} we will adapt the mathematical results of~\cite{Claeys} to determine an exact interpolation formula for the EFP in the limit of large $L$ in the quantum Ising chain ($\gamma=1$); in section~\ref{sec:expansion} we will present the first terms of a series expansion of the $\tau$ function relaying on the recent irregular~\cite{Nagoya} conformal block representation proposed in~\cite{Gamayun0, Gamayun, Lisovyy}. Our results will be extended to $\gamma\not=1$  in section~\ref{sec:results}, through a conjecture, and  thoroughly tested against numerical lattice calculations. The agreement is excellent. Finally in section~\ref{sec:xx}  we will focus on the case $\gamma=0$. Recalling that  the EFP is the Fredolhm determinant of the so-called sine kernel~\cite{Dyson, Jimbo2}, for completeness, we rediscuss how its exact asymptotic Gaussian behaviour~\cite{Shiroishi} can be determined  in a double-scaling limit. Two appendices complete the paper.

\section{Emptiness formation probability in the
  ground state of the XY spin chain}\label{sec:efp}

Consider the anisotropic XY spin-1/2 chain in a transverse
magnetic field with Hamiltonian given in (\ref{xy_chain}).
The model reduces to the quantum Ising chain for $\gamma=1$ 
while corresponds to free fermion if $\gamma=0$, that is the XX spin
chain. We will always consider periodic boundary conditions, $\sigma_{n+N}^\mu
=\sigma_n^\mu$.

We are interested in the analysis of the EFP in the ground state $\ket{{\rm GS}}$ of the
XY spin chain (\ref{xy_chain}), which can be defined as
\begin{equation*}\label{efp}
  \mathcal{P}(L, h, \gamma)=\bra{{\rm GS}} \prod_{l=1}^L
  \frac{1-\sigma_l^z}{2}\ket{{\rm GS}}.
\end{equation*}
As we anticipated in the introduction, this is the probability 
to find a string of $L$ consecutive down spins in the state 
$\ket{{\rm GS}}$.

The Hamiltonian (\ref{xy_chain}) can be recast in a quadratic fermionic form by introducing 
the Jordan-Wigner transformation,
\begin{equation*}
  \psi_n\equiv\prod_{j=1}^{n-1}(-\sigma_j^z)
  \frac{\sigma_n^x-{\rm i}\sigma_n^y}{2},
  \quad
  \sigma_n^z=2\psi_n^\dagger \psi_n-1,
\end{equation*}
and rewriting the spin operators in terms of the
creation/annihilation fermionic operators  
$\psi_n^\dagger$, $\psi_n$; one obtains 
\begin{equation}\label{fermionic_chain}
  H=Nh+\sum_{n=1}^{N}\left[\gamma(\psi_n^\dagger \psi_{n+1}^\dagger
    -\psi_n\psi_{n+1})+\psi_n^\dagger\psi_{n+1}
    +\psi_{n+1}^\dagger \psi_n-2h\psi_n^\dagger \psi_n\right].
\end{equation}
The boundary conditions for the fermionic operators are assumed periodic $\psi_{n+N}=\psi_n$,
although this choice is irrelevant in the thermodynamic limit $N\rightarrow\infty$
\footnote{In principle, after applying
  the Jordan-Wigner transformation, we must split the Hilbert space
  into two sectors. For the states with an odd number of particles
  the resulting fermionic chain is periodic (Ramond sector) while for
  those with an even number the chain is antiperiodic (Neveu-Schwarz sector).
  In the observables, the difference between both sectors are terms that
  go to zero in the thermodynamic limit. Since in this work we will restrict
  to this limit, they can be neglected and assume periodic boundary conditions
  for simplicity.}. In terms of the spinless fermions, the EFP
can be expressed as the expectation value
\begin{equation*}
  \mathcal{P}(L, h, \gamma)=\langle\prod_{l=1}^L\psi_l\psi_l^\dagger\rangle
\end{equation*}
over the ground state $\ket{{\rm GS}}$.
As it was shown in \cite{Shiroishi} and
\cite{Abanov, Franchini}, using the Wick theorem,
$\mathcal{P}(L, h, \gamma)$ can be written as the
determinant
\begin{equation}\label{efp_det}
  \mathcal{P}(L, h, \gamma)=|\det S|,
\end{equation}
where $S$ is the $L\times L$ matrix
with entries
\begin{equation*}
  S_{nm}=\langle\psi_n\psi_m^\dagger\rangle+
  \langle \psi_n^\dagger \psi_m^\dagger\rangle,
  \quad 1\leq n,m\leq L,
\end{equation*}
built from the two-point correlation
functions restricted to the interval of length $L$.
In the thermodynamic limit, $S$
becomes the Toeplitz matrix
\begin{equation}\label{matrix_s}
  S_{nm}=\frac{1}{2\pi}
  \int_{0}^{2\pi} g(\theta){\rm e}^{\ii\theta(n-m)}
      {\rm d}\theta,
\end{equation}
generated by the symbol \cite{Shiroishi, Abanov, Franchini}
\begin{equation}\label{symbol_g}
  g(\theta)=\frac{1}{2}+\frac{\cos\theta-h+\ii\gamma\sin\theta}
  {2\sqrt{(\cos\theta-h)^2+\gamma^2\sin^2\theta}}.
\end{equation}

\begin{figure}[t]
  \centering
  \includegraphics[width=0.4\textwidth]{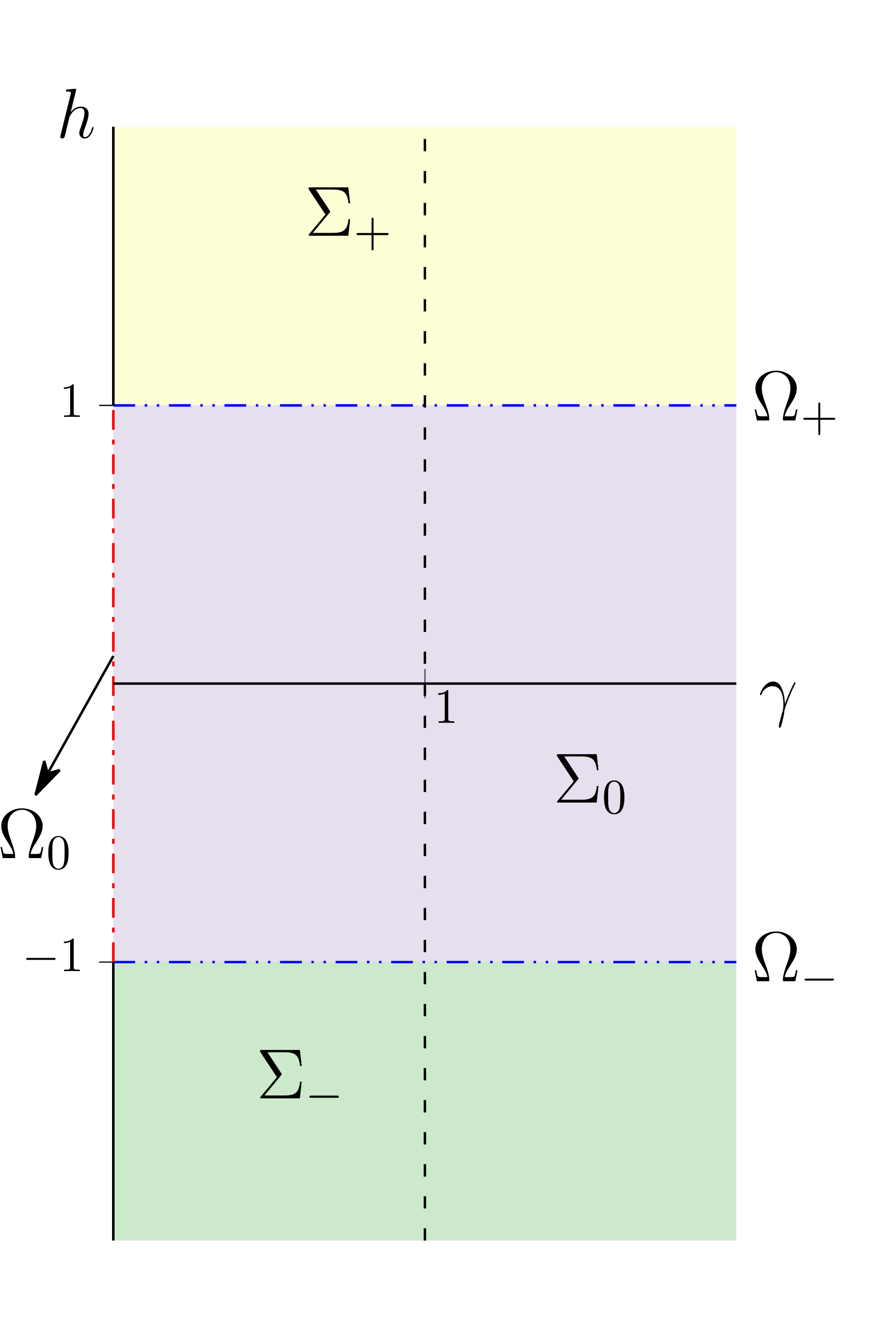}
  \caption{Plane of parameters $(\gamma, h)$ of the XY spin chain (\ref{xy_chain})
    divided in regions depending on the asymptotic behaviour of the EFP. The chain is non-critical in regions $\Sigma_\pm$ and $\Sigma_0$ while it
    is critical along the lines $\Omega_\pm$ and $\Omega_0$. }
  \label{phase_diagram_xy}
   \end{figure}

In~\cite{Abanov, Franchini}, the asymptotic large-$L$ behaviour of $\mathcal{P}$  was studied. 
The authors made use of the Szeg\H{o} theorem \cite{Szego}, the Fisher-Hartwig conjecture \cite{Fisher, Basor1}
and its generalization \cite{Basor2, Basor3}. These theorems/conjectures,
that we review in the appendix \ref{sec:appendixa}, determine the asymptotic large-$L$ behaviour of the Toeplitz
determinant (\ref{efp_det}). It will be worth to rediscuss and complement here slightly
the main results of \cite{Abanov, Franchini}. In those works, several regions in the parameter space $(\gamma, h)$
were distinguished depending on the asymptotic behaviour of the EFP. We depict
them in fig. \ref{phase_diagram_xy} for future reference.

  \subsection{Off-critical Region $\Sigma_-$: $\gamma\neq 0$ and $h<-1$.}
    
    In this region, the symbol does not have any
    Fisher-Hartwig singularity, that is, $g(\theta)$
    has neither zeros or jump discontinuities and it
    satisfies the smoothness condition (\ref{smoothness_cond}).
    Therefore, the Szeg\H{o}
    theorem (\ref{szego}) applies and it follows
    \begin{equation}\label{asymp_sigma_-}
      \log \mathcal{P}(L, h, \gamma)= A(h, \gamma) L+ E(h, \gamma)+o(1).
    \end{equation}
    To compute the coefficients $A$ and $E$ in \eqref{asymp_sigma_-}, it will be convenient to rewrite the symbol $g(\theta)$ of
the Toeplitz matrix (\ref{matrix_s}) in the following form.
Let us introduce the Laurent polynomials
\begin{equation}\label{laurent}
  \Phi(z)=\frac{z}{2}-h+\frac{z^{-1}}{2},\quad
  \Xi(z)=\frac{\gamma}{2}\left(z-z^{-1}\right),
\end{equation}
then (note that $g(\theta)=\mathcal{G}(\e^{\ii \theta})$)
\begin{equation*}
  \mathcal{G}(z)=\frac{1}{2}
  \left(1+\frac{\Phi(z)+\Xi(z)}
       {\sqrt{\Phi(z)^2-\Xi(z)^2}}\right).
\end{equation*}

We then decompose  $\mathcal{G}(z)$ in polar form $\mathcal{G}(z)=|\mathcal{G}(z)|\e^{\ii\arg \mathcal{G}(z)}$;  for $z$ on the unit circle, it turns out  
$$|\mathcal{G}(z)|^2=\frac{1}{2}\left(1
+\frac{\Phi(z)}{\sqrt{\Phi(z)^2-\Xi(z)^2}}\right), \quad {\rm and} 
\quad \arg \mathcal{G}(z)=\log\left(\frac{\Phi(z)-\Xi(z)}
{\Phi(z)+\Xi(z)}\right)^{\ii/4}.$$
Finally, we find that
\begin{equation}\label{symbol_G}
  \mathcal{G}(z)=V(z)
  \left[\frac{(z-z_+)(z-z_-)}{(zz_+-1)(zz_--1)}\right]^{1/4}
\end{equation}
with
\begin{equation}\label{branch_points}
  V(z)=\sqrt{\frac{1}{2}+
\frac{\Phi(z)}{2\sqrt{\Phi(z)^2-\Xi(z)^2}}},\quad\text{and}\quad z_{\pm}=\frac{h\pm\sqrt{h^2+\gamma^2-1}}{1+\gamma}.
\end{equation}
The expression in~\eqref{symbol_G} for the symbol can be continued analytically  
from the unit circle 
to the Riemann sphere $\bar{\mathbb{C}}=\mathbb{C}\cup \{\infty\}$. Note that $z_+$, $z_-$, $z_+^{-1}$, and $z_-^{-1}$
are branch points of the function $\mathcal{G}(z)$. Although in this section we are focussing on the case $h<-1$, it is useful to describe the analyticity properties of $\mathcal{G}(z)$ in ~\eqref{symbol_G} more in general. Observe that for $h^2+\gamma^2<1$, the branch points  are complex conjugated (i.e. $\bar{z}_+=z_-$), while for $h^2+\gamma^2\geq 1$,
they are real. Outside the critical lines, two of them
are inside the unit disk while the other pair is outside.
Thus we will always take as branch cuts for $\mathcal{G}(z)$
two curves, each one joining the couple of branch points
located at the same side of the unit circle without
intersecting the latter. For instance for $\gamma=1$ and $h<-1$ the branch cuts can be taken as $(-\infty,h]\cup[h^{-1},0]$.
On the other hand, along the
critical lines $\Omega_\pm$, where $|h|\rightarrow 1$, two of the branch points
collide on the unit circle. Hence the two branch cuts
join together in a single cut that crosses the unit circle, giving rise
to a jump discontinuity in $g(\theta)$, see for instance fig. \ref{fig:scaling_limit_sigma_-}. 

Taking into account the factorization (\ref{symbol_G}) and the definition of the branch cuts given earlier, we can expand the logarithm of the symbol in Fourier series as (see appendix~\ref{sec:appendixa})
\begin{equation}
\label{FourierG}
 \log\mathcal{G}(z)=(\log V)_0+\sum_{k>0}z^k\left((\log V)_k-\frac{z_-^{-k}-z_+^+}{4k}\right)+\sum_{k>0}z^{-k}\left((\log V)_{k}+\frac{z_{-}^{-k}-z_{+}^{k}}{4k}\right),
\end{equation}
and
    $$(\log V)_{\pm k}=\frac{1}{2\pi}
    \int_0^{2\pi}\log V(\e^{\ii \theta})\e^{\pm \ii\theta k}{\rm d}\theta.$$

It then follows from~\eqref{FourierG} that the coefficient $A(h, \gamma)$ in~\eqref{asymp_sigma_-} is 
    equal to the zero mode of the function $\log V(\e^{\ii\theta})$,
    \begin{eqnarray*}
    A(h,\gamma)=(\log V)_0&=&\frac{1}{2\pi}
    \int_0^{2\pi}\log V(\e^{\ii\theta}){\rm d}\theta \\
    &=&\frac{1}{4\pi}
    \int_0^{2\pi}\log\left[\frac{1}{2}+\frac{-h+\cos\theta}
    {2\sqrt{(-h+\cos\theta)^2+\gamma^2\sin^2\theta}}\right]
    {\rm d}\theta,
    \end{eqnarray*}
    while the $O(1)$ term, $E(h,\gamma)$, can be written as
    $$E(h, \gamma)=E[V]-\frac{1}{16}\log\left(1-\frac{\gamma^2}
    {1-h^2}\right),$$
    where $E[\bullet]$ is defined in the appendix \ref{sec:appendixa}.
    
    Note that, in the limit $h\to -\infty$, the ground state is 
    $\ket{\downarrow \downarrow \cdots \downarrow}_z$ and $\mathcal{P}=1$.
    Indeed, it is easy to verify that when $h\to-\infty$, the function 
    $\log V(\ee^{\ii\theta})\to 1$, and $A(h,\gamma), E(h, \gamma)\to 0$.

  \subsection{Off-critical Region $\Sigma_0$: $\gamma\neq 0$ and $|h|<1$.}

    In this case, the symbol $g(\theta)$ presents a
    Fisher-Hartwig singularity at $\theta=\pi$: The
    modulus of $g(\theta)$ vanishes and its argument
    has a jump discontinuity.
    
    Hence one needs
    to apply the Fisher-Hartwig conjecture to obtain the
    asymptotic large-$L$ behaviour of $\mathcal{P}$. The symbol admits a
    unique factorization of the form (\ref{fh_fact}),
    \begin{equation}\label{g_sigma_0}
    g(\theta)=W^0(\theta)
    \left[\frac{(1-z_+\e^{-\ii\theta})(1-z_-\e^{-\ii\theta})}
      {(\e^{\ii\theta}z_+-1)(\e^{\ii\theta}z_--1)}\right]^{1/4}
    [2-2\cos(\theta-\pi)]^{1/2}{\rm e}^{\ii/2[\theta-\pi-\pi\sign(\theta-\pi)]}
    \end{equation}
    where
    $$W^0(\theta)=V(\e^{\ii\theta})[2-2\cos(\theta-\pi)]^{-1/2}.$$
    Therefore, according to (\ref{fh}), the EFP behaves as
    \begin{equation}\label{asymp_sigma_0}
      \log \mathcal{P}(L, h, \gamma)= A(h,\gamma)L+E(h, \gamma)+o(1).
    \end{equation}
    The coefficient $A(h,\gamma)$ is now the zero mode of $\log W^0$,
    $$A(h, \gamma)=(\log W^0)_0=\frac{1}{4\pi}
    \int_0^{2\pi}\log\left[\frac{1}{2}+\frac{-h+\cos\theta}
    {2\sqrt{(-h+\cos\theta)^2+\gamma^2\sin^2\theta}}\right]
    {\rm d}\theta.$$
    The term $E(h,\gamma)$ instead reads
    $$E(h, \gamma)=E[W^0]-W_-^0(\pi)+\frac{1}{16}\log\frac{1-h^2}{\gamma^2}
    -\frac{1}{4}\log\frac{1+h}{\gamma},$$
    with
    $$W_\pm^0(\theta)=\sum_{k=1}^\infty(\log W^0)_{\pm k}\e^{\ii\theta k}$$
    and $E[\bullet]$, the sum over Fourier coefficients defined in appendix~\ref{sec:appendixa}.

    For the values $h=0$, $\gamma=1$, the ground state of the XY spin
    chain is degenerate; a basis for the degenerate subspace is given by fully polarized states in the $x$ direction, that is
    $$
    \left\{(\ket{\rightarrow \rightarrow \cdots \rightarrow}_x, \ket{\leftarrow \leftarrow \cdots \leftarrow}_x\right\}.$$
    It is straightforward to see that for these states, 
    the EFP is $\mathcal{P}=(1/2)^L$. Note that the same
    is obtained from the asymptotic expression (\ref{asymp_sigma_0}),
    which in this case is exact since $A(0, 1)=-\log 2$ and $E(0, 1)=0$. This observation was extended in~\cite{Franchini} to the ground state along the circle $h^2+\gamma^2=1$.
    
    \subsection{Off-critical Region $\Sigma_+$: $\gamma\neq 0$ and $h>1$.}

    The symbol $g(\theta)$ vanishes and its argument has a jump discontinuity
    at the points $\theta=0$ and $\theta=\pi$: Therefore, it has two Fisher-Hartwig singularities.
    In the region $\Sigma_+$, there are two possible factorizations (\ref{gfh_fact}):
    $$g(\theta)=W^+(\theta)
    (2-2\cos\theta)^{1/2}\e^{-\ii/2(\theta-\pi\sign\theta)}
    (2-2\cos(\theta-\pi))^{1/2}
    \e^{\ii/2(\theta-\pi-\pi\sign(\theta-\pi))},$$
    and
    $$g(\theta)=\tilde{W}^+(\theta)
    (2-2\cos\theta)^{1/2}\e^{\ii/2(\theta-\pi\sign\theta)}
    (2-2\cos(\theta-\pi))^{1/2}\e^{-\ii/2(\theta-\pi-\pi\sign(\theta-\pi))}.$$
    In order to determine the asymptotic behaviour of $\mathcal{P}$,
    we must apply the generalized Fisher-Hartwig conjecture (\ref{gfh}),
    obtaining
    $$\log \mathcal{P}(L, h, \gamma)= A(h, \gamma) L+
    \log\left[E_1(h, \gamma)+(-1)^L E_2(h,\gamma)\right]+o(1).$$
    In this case,
    $$A(h,\gamma)=(\log W^+)_0=(\log \tilde{W}^+)_0,$$
    and
    $$E_1(h, \gamma)=\ee^{E[W^+]-W_+^+(0)-W_-^+(\pi)},
    \quad E_2(h,\gamma)=\ee^{E[W^+]-W_+^+(\pi)-W_-^+(0)}.$$
    When $h\to\infty$, the ground state of the chain is $\ket{\uparrow
    \uparrow \cdots \uparrow }_z$ and, consequently, $\mathcal{P}=0$. Indeed,
    in this limit, $\log V(\ee^{\ii\theta})\to -\infty$ and
    the coefficient $A(h,\gamma)\to -\infty$.
    
The off-critical regions which we have discussed so far are separated by the critical lines $\Omega_\pm$. The free fermion line $\Omega_0$ has some peculiar properties that we will discuss at the end of this section.

\subsection{Critical line $\Omega_-$: $\gamma\neq 0$ and $h=-1$.}
  
    Along the line $\Omega_-$, the symbol has a Fisher-Hartwig singularity
    at $\theta=\pi$. For this value of $\theta$, its argument is discontinuous.
    The symbol admits different factorizations of the form (\ref{gfh_fact}).
    Nevertheless, in this case, the contribution of such ambiguity
    to the EFP is a subleading $o(1)$ term. The leading terms can actually
    be computed taking the factorization
    $$g(\theta)=V(\e^{\ii\theta})\left(\frac{\e^{\ii\theta}-z_+}{\e^{\ii\theta}z_+-1}\right)^{1/4}
    \e^{\ii/4(\theta-\pi-\pi\sign(\theta-\pi))},$$
    and applying the Fisher-Hartwig conjecture (\ref{fh}). One 
    obtains 
    \begin{equation}\label{asymp_omega_-}
      \log \mathcal{P}(L, -1,\gamma)=
      A(-1, \gamma)L-\frac{1}{16}\log L+E(-1, \gamma)+o(1),
    \end{equation}
    where the coefficient of the linear term is now
    $$A(-1, \gamma)=(\log V)_0,$$
    while the $L$-independent term $E(-1, \gamma)$ can be computed from
    $$E(-1, \gamma)=E[V]-\frac{1}{16}\log\gamma+\log\left[G\left(\frac{3}{4}\right)
      G\left(\frac{5}{4}\right)\right],$$
    and $G$ denotes the Barnes $G$ function.
      
  The coefficient of the logarithmic term in (\ref{asymp_omega_-}) has a field theoretical interpretation. Indeed, in~\cite{Stephan},  the EFP along the critical line $\Omega_{-}$ was analysed 
    using CFT methods. In CFT, the EFP can be expressed as the normalized
    free energy of an infinite cylinder with a slit of length $L$, along which free boundary conditions are imposed. From
    the CFT analysis, one concludes  that the coefficient of the logarithmic
    term in (\ref{asymp_omega_-}) is $c/8$ where $c$ is the central charge of the CFT which describes the low-energy physics. Taking
    into account that the line $\Omega_-$ belongs to the Ising universality
    class, with $c=1/2$, the coefficient $1/16$ in \eqref{asymp_omega_-} follows. 
    
  \subsection{Critical line $\Omega_+$: $\gamma\neq0$ and $h=1$.}

    Here the symbol has two Fisher-Hartwig singularities
    at $\theta=0$ and at $\theta=\pi$. The singularity at 
    $\theta=0$ is a jump discontinuity in the argument while 
    the singularity at $\theta=\pi$ is a combination of a zero
    and a jump discontinuity. There are different possibilities (\ref{gfh_fact}) of factorizing
    the symbol, however, as it also occurs in the critical line
    $\Omega_-$, the leading terms can be obtained choosing the
    particular factorization
    $$g(\theta)=W^0(\theta)\left(\frac{1-z_-\e^{-\ii\theta}}{\e^{\ii\theta}z_--1}\right)^{1/4}
    \e^{\ii/4(\theta-\pi\sign\theta)}
    (2-2\cos(\theta-\pi))^{1/2}\e^{\ii/2(\theta-\pi-\pi\sign(\theta-\pi))}.$$
    The Fisher-Hartwig conjecture (\ref{fh}) predicts in this case that
    \begin{equation}\label{asymp_omega_+}
      \log \mathcal{P}(L, 1, \gamma)=A(1, \gamma)L-\frac{1}{16}\log L+E(1, h)+o(1),
    \end{equation}
    with
    $$A(1, \gamma)=(\log W^0)_0,$$
    and
    $$E(1, h)=E[W^0]-W_-^0(\pi)-\frac{1}{4}\log 2+\frac{1}{16}\log\gamma^3+
    \log\left[G\left(\frac{3}{4}\right)G\left(\frac{5}{4}\right)\right].$$
    
    For $h=1$ ($\gamma\neq 0$), one can also study the EFP by employing CFT methods.
    As it happens on the line $\Omega_-$, the EFP corresponds to the
    free energy of an infinite cylinder with a slit of length $L$ along which 
    fixed boundary conditions are now imposed. 
    This difference in the boundary conditions does not affect the logarithmic
    term in (\ref{asymp_omega_+}), whose coefficient is still $c/8$, but it can 
    modify the subleading terms \cite{Stephan}.

    \subsection{Critical line $\Omega_0$: $\gamma=0$, $|h|<1$.}

    The line $\Omega_0$ corresponds to the critical XX spin chain.  
    For $\gamma=0$, the symbol $g(\theta)$ is the
    piecewise constant function~
    \begin{equation}\label{symbol_xx}
      g(\theta)=\left\{\begin{array}{ll} 1, & k_F \leq \theta \leq 2\pi-k_F,\\
      0, & 0<\theta < k_F,\quad \mbox{or}\quad 2\pi-k_F< \theta < 2\pi,
      \end{array}\right.
    \end{equation}
    where $k_F=\arccos h$ is the Fermi momentum. The symbol (\ref{symbol_xx}) can be interpreted as the 
    density of occupied modes in the ground state. Such a state is a Dirac sea filled by all the particles with negative energy, i.e. with momentum between $k_F$ and $2\pi-k_F$. The low energy field theory is 
    a massless Dirac fermion with $c=1$, or equivalently a Luttinger liquid. 

    For symbols with compact support as in (\ref{symbol_xx}), the
    asymptotics of the corresponding determinant is derived
    from the Widom theorem \cite{Widom}, see appendix \ref{sec:appendixa}. In particular, denoting 
    $\mathcal{P}_0(L, k_F)$ the EFP at $\Omega_0$, one obtains that
     \begin{equation}\label{efp_omega_0}
      \log \mathcal{P}_0(L, k_F)= \frac{L^2}{2}\log\left(1-\sin^2\frac{k_F}{2}\right)-
      \frac{1}{4}\log\left(L\sin\frac{k_F}{2}\right)+\log[\sqrt{\pi}G(1/2)^2]+o(1).
     \end{equation}
    This result  was firstly pointed out in~\cite{Shiroishi}.
    As recalled in the introduction, the asymptotic behaviour of the 
    EFP is radically different from that characterizing the critical 
    lines $\Omega_\pm$: The leading term in (\ref{efp_omega_0}) 
    is  $O(L^2)$ rather  than $O(L)$ as in 
    (\ref{asymp_omega_-}) and (\ref{asymp_omega_+}). Furthermore, 
    the prefactor of the logarithmic term is no longer interpretable as $c/8$,
    although  speculations about its universality have been put forward in \cite{Stephan}.
    
When $|h|>1$, $g(\theta)=0$, the mass gap is non-zero and the ground
state of the chain is the Fock vacuum. Therefore, it is straightforward to see
that $\mathcal{P}_0=1$ for $h<-1$ ($\ket{{\rm GS}}=\ket{\downarrow \downarrow
  \cdots \downarrow}_z$) while $\mathcal{P}_0=0$ for $h>1$ ($\ket{{\rm GS}}=
\ket{{\uparrow \uparrow \cdots \uparrow}}_z$).

\section{Transition from the off-critical regions to
  the critical lines}\label{sec:transition}

The purpose of this paper is to analyze the behaviour of
the EFP approaching a quantum critical point. More specifically, we will discuss 
how the EFP interpolates from the asymptotic regime
observed in the off-critical regions $\Sigma_-$ and $\Sigma_0$,
see (\ref{asymp_sigma_-}) and (\ref{asymp_sigma_0}), and the one that follows
along the critical lines $\Omega_-$ and $\Omega_+$ respectively, see (\ref{asymp_omega_-})
and (\ref{asymp_omega_+}). As we have reviewed in the previous section, in the critical
regions, the logarithm of the EFP, which is $O(L)$, develops also subleading logarithmic corrections 
$O(\log L)$. In mathematical terms, the approach towards the critical lines 
is described by a double-scaling limit in which the length of
the interval $L$ diverges while the inverse of the correlation length, proportional to $\log|h|$, vanishes.

\begin{description}
  \item \textit{Transition from the region $\Sigma_-$ to the line $\Omega_-$}

In the region $\Sigma_-$, the symbol $g(\theta)$ does not present any
Fisher-Hartwig singularity and it is analytic on the unit circle.
In this case, the branch points $z_+$ and $z_{-}^{-1}$ of $\mathcal{G}(z)$, see (\ref{symbol_G}), are real
and they are inside the unit disk while their inverses are outside.
As depicted in fig.~\ref{fig:scaling_limit_sigma_-}, when approaching the line $\Omega_-$, i.e. when $h\to-1^-$, the branch
points $z_{-}$ and $z_{-}^{-1}$ move towards $z=-1$ and they eventually intersect the unit circle for
$h=-1$. Such a  merging of  singularities gives rise to the jump discontinuity of the symbol $g(\theta)$
at $\theta=\pi$ along the line $\Omega_-$. 

\begin{figure}[t]
  \centering
  \includegraphics[width=0.8\textwidth]{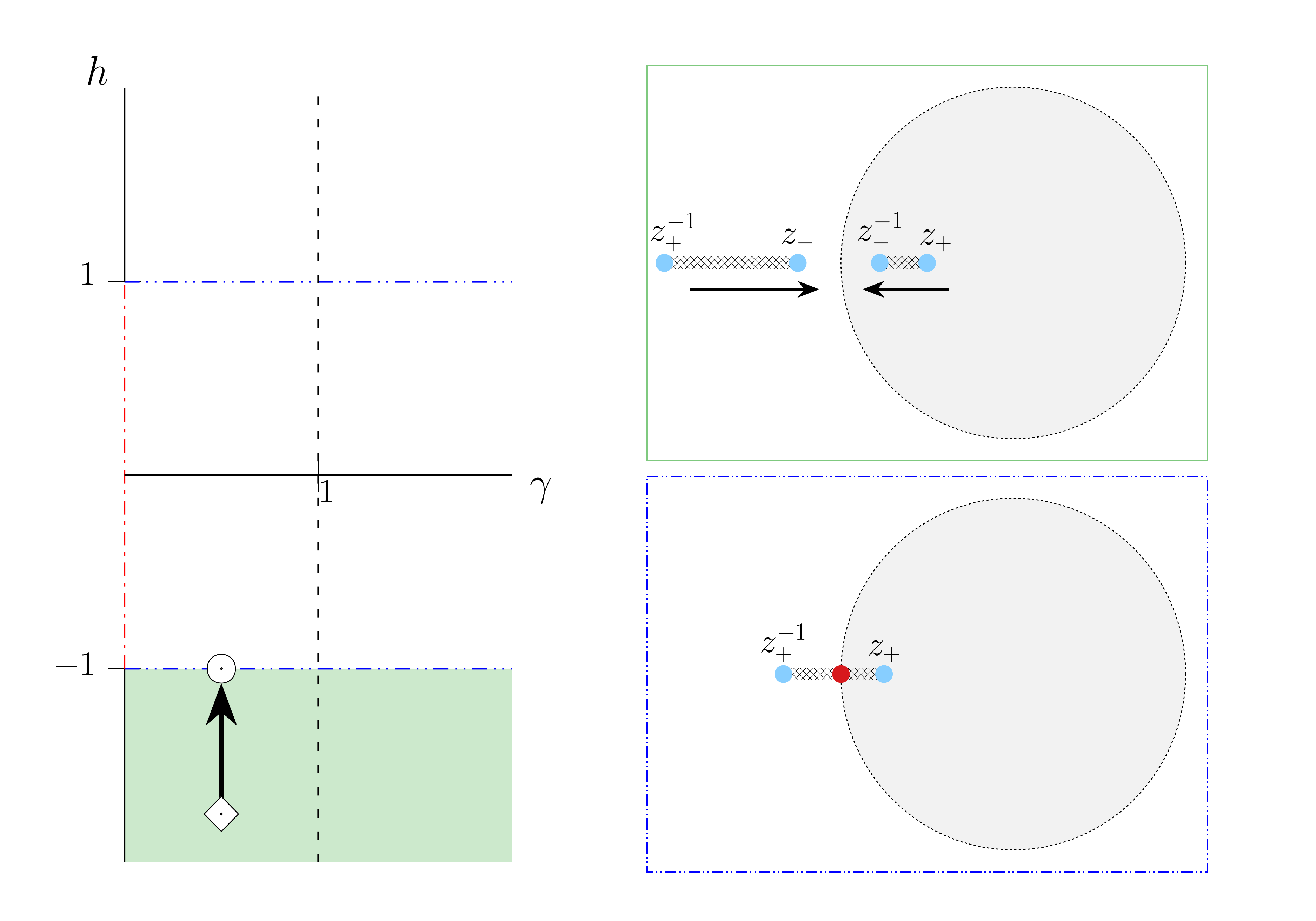}
  \caption{For a chain with $h<-1$, indicated by
    $\mydiamond$ in the plane $(\gamma, h)$ on the left hand side, the branch points
    $z_+$ and $z_-^{-1}$  of $\mathcal{G}(z)$ are inside the unit disk while
    their inverses are outside, as represented in the upper
    right panel. If $h\to -1^-$, with $\gamma$ fixed, then
    $z_-\to z_-^{-1}$. When approaching the critical line $h=-1$,
    denoted by the point $\mycircle$, the
    branch points $z_-$ and $z_-^{-1}$ merge at $z=-1$, as shown in
    the lower right panel. The resulting branch cut joining $z_+$ and $z_+^{-1}$
    crosses the unit circle, producing a jump discontinuity in the symbol
    $g(\theta)$ of the Toeplitz matrix $S$.
 }
  \label{fig:scaling_limit_sigma_-}
   \end{figure}

  \item \textit{Transition from the region $\Sigma_0$ to the line $\Omega_+$}

In the region $\Sigma_0$, the symbol $g(\theta)$ has already a Fisher-Hartwig
singularity that it is a combination of a zero and a jump discontinuity
at $\theta=\pi$, see (\ref{g_sigma_0}). This singularity is also present
along the critical line $\Omega_+$. Along this line, however, the symbol has another
singularity at $\theta=0$, which is a jump. As
fig. \ref{fig:scaling_limit_sigma_0} illustrates, this discontinuity is produced
by a similar mechanism that occurs also in the transition from the region $\Sigma_-$ to $\Omega_-$.
In this case, as $h\to 1^-$, the branch points $z_+$ and $z_+^{-1}$ come together
at $z=1$, producing the jump at $\theta=0$ in $g(\theta)$. In the coalescence limit,
the Fisher-Hartwig singularity at $\theta=\pi$ is not affected.

\begin{figure}[t]
  \centering
  \includegraphics[width=0.8\textwidth]{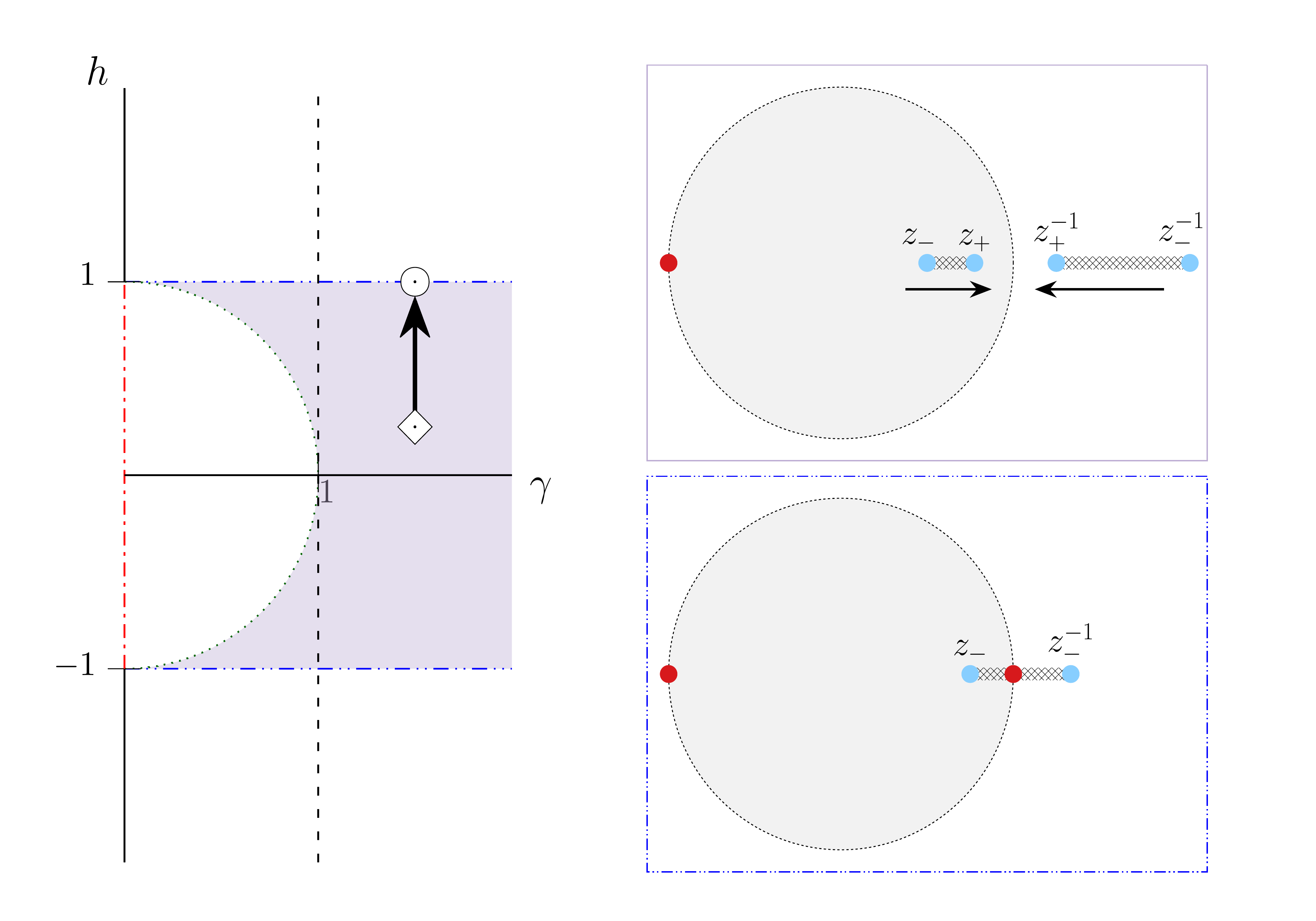}
  \caption{Consider a chain with $|h|<1$ and $h^2+\gamma^2>1$ and identified
    by the point $\mydiamond$ in the
    plane $(\gamma, h)$ on the left hand side.
    For this theory, the branch points $z_-$ and $z_+$ of $\mathcal{G}$
    are inside the unit disk as we depict in the upper right panel.
    The symbol has a Fisher-Hartwig singularity
    at $z=-1$, represented by the dot $\textcolor{myred}{\mycirclefilled}$.
    If $h\to 1^-$ and $\gamma$ fixed, then $z_+\to z_+^{-1}$.
    As it is illustrated in the lower right panel, at
    the point $\mycircle$ along the critical line
    $\Omega_+$, they merge on the unit
    circle, $z_+=z_+^{-1}=1$, giving rise to a second Fisher-Hartwig singularity,
    a jump, in the symbol $g(\theta)$ of $S$. The singularity at $z=-1$ remains
    untouched in the transition from the region $\Sigma_0$ to the critical line $\Omega_+$.}
  \label{fig:scaling_limit_sigma_0}
\end{figure}
\end{description}

In the next section, we introduce the mathematical
tools which are needed to study the behaviour
of a Toeplitz determinant when a Fisher-Hartwig singularity emerges 
from the collision of two branch points on the unit circle.

\section{Interpolation between Fisher-Hartwig 
         and Szeg\H{o} asymptotics: Toeplitz meets Painlev\'e}\label{sec:interpolation}

Ref. \cite{Claeys} studied the transition between the Szeg\H{o} and the 
Fisher-Hartwig regimes when a single singularity emerges in a regular symbol. This is
the case of the transition between the region $\Sigma_-$ and the line
$\Omega_-$.
More specifically, in~\cite{Claeys}, such a transition is analyzed
for the Toeplitz determinant generated by the particular symbol
\begin{equation}\label{ick_symbol}
  f_t(z)=(z-{\rm e}^t)^{\alpha+\beta}(z-{\rm e}^{-t})^{\alpha-\beta}
  z^{-\alpha+\beta}\e^{-\ii\pi(\alpha+\beta)}{\rm e}^{Q(z)},
\end{equation}
where $t\geq 0$, $\alpha,\beta\in\mathbb{C}$ with
${\rm Re}\,\alpha>-1/2$ and $\alpha\pm\beta\neq -1,-2,\dots$.
The function $Q(z)$ is analytic on the unit circle. Notice that,
in the region $\Sigma_-$, the symbol (\ref{symbol_G}) for the quantum Ising chain
($\gamma=1$) is of the form (\ref{ick_symbol}) with
$\alpha=0$, $\beta=1/4$, $t=\log|h|$, and
$Q(z)=\log((-h)^{-1/4} V(z))+\ii\pi/4$. As discussed in~\cite{Claeys}, the symbol for the two-point function of the magnetization in the two-dimensional Ising model is also of the form~(\ref{ick_symbol}) with $\alpha=0$ and $\beta=-1/2$.

For $t>0$ the symbol (\ref{ick_symbol}) has not any Fisher-Hartwig singularity
and the Szeg\H{o} theorem gives the asymptotic large-$L$ behaviour of the
determinant $D_L[f_t]$. Considering the
Fourier expansion of the function $Q(z)$,
\begin{equation*}
  Q(z)=\sum_{k\in\mathbb{Z}}Q_kz^k,
\end{equation*}
the Szeg\H{o} theorem (\ref{szego}) predicts for the particular
symbol (\ref{ick_symbol})
\begin{equation}\label{szego_ft}
  \log D_L[f_t]=s_0L+\sum_{k=1}^\infty k s_k s_{-k}+o(1),
\end{equation}
where the $s_k$'s are the Fourier modes of $\log f_t$,
\begin{equation*}
  s_0=Q_0+(\alpha+\beta)t,\quad s_{\pm k}=Q_{\pm k}-(\alpha\pm \beta)\frac{\e^{-tk}}{k}.
\end{equation*}
On the other hand, for $t=0$, the symbol $f_t({\rm e}^{{\rm i}\theta})$ has a
Fisher-Hartwig singularity at $\theta=0$. This singularity is a combination
of a jump (for $\alpha=0$ and $\beta\neq 0$) and a zero
(for $\alpha\neq 0$ and $\beta=0$). In this case, the Fisher-Hartwig
conjecture (\ref{fh}) gives the asymptotic behaviour of $D_L[f_0]$, 
\begin{equation}\label{f-h_ft}
  \log D_L[f_0]= Q_0L+(\alpha^2-\beta^2)\log L+E+o(1),
\end{equation}
where
\begin{equation*}
  E=\sum_{k=1}^\infty\left(k Q_k Q_{-k}-(\alpha-\beta)Q_k-(\alpha+\beta)Q_{-k}\right)+
  \log\frac{G(1+\alpha+\beta)G(1+\alpha-\beta)}{G(1+2\alpha)}.
\end{equation*}

In Theorem 1.1 of~\cite{Claeys} was derived an asymptotic expansion 
for $D_L[f_t]$ that interpolates between the Szeg\H{o} (\ref{szego_ft}) 
and the Fisher-Hartwig (\ref{f-h_ft}) asymptotics. In particular, the theorem states that for $L\to\infty$,
\begin{multline}\label{ick_asymptotics}
  \log D_L[f_t]=s_0L+\sum_{k=1}^\infty k s_k s_{-k}+(\alpha^2-\beta^2)\log(2tL)
  \\+\tilde{\Omega}(2tL)+
  \log \frac{G(1+\alpha+\beta)G(1+\alpha-\beta)}{G(1+2\alpha)}+o(1),
\end{multline}  
where $o(1)$ is uniform for $0<t<t_0$ with $t_0$ small enough. The 
term $\tilde{\Omega}(2tL)$ is of the form
\begin{equation}\label{omega_tilde}
  \tilde{\Omega}(2tL)=\int_0^{2tL}\frac{\zeta(x)-\alpha^2+\beta^2}{x}{\rm d}x,
\end{equation}
and $\zeta(x)$ is the particular solution of the Jimbo-Miwa-Okamoto
form of the Painlev\'e V equation~\cite{Jimbo}
\begin{equation}\label{pV_its}
  (x \zeta'')^2=(\zeta-x\zeta'+2(\zeta')^2+2\alpha\zeta')^2-
  4(\zeta')^2(\zeta'+\alpha+\beta)(\zeta'+\alpha-\beta),
\end{equation}
satisfying the boundary conditions
\begin{equation}\label{asymptotics_sigma_t_small}
  \zeta(x)\sim \alpha^2-\beta^2
  -\frac{\alpha^2-\beta^2}{2\alpha}\left(x-C_{\alpha,\beta}x^{1+2\alpha}\right),
  \quad \mbox{for} \quad x\to 0,\,\, 2\alpha\not\in\mathbb{Z},
\end{equation}
and
\begin{equation}
\label{asymptotics_sigma_t_large}
  \zeta(x)\sim -\frac{1}{\Gamma(\alpha-\beta)
    \Gamma(\alpha+\beta)}x^{-1+2\alpha}\e^{-x},
  \quad \mbox{for}\quad x\to\infty,
\end{equation}
with
\begin{equation}\label{c_ab}
  C_{\alpha,\beta}=\frac{\Gamma(1+\alpha+\beta)\Gamma(1+\alpha-\beta)}
  {\Gamma(1-\alpha+\beta)\Gamma(1-\alpha-\beta)}\cdot
  \frac{\Gamma(1-2\alpha)}{\Gamma(1+2\alpha)^2}
  \cdot\frac{1}{1+2\alpha}.
\end{equation}
 Remarkably, from eq.~\eqref{ick_asymptotics} and eqs.~\eqref{asymptotics_sigma_t_small}-\eqref{asymptotics_sigma_t_large}, one can recover~\eqref{f-h_ft} in the limit $t\rightarrow 0$ while sending $t\rightarrow\infty$ one obtains~\eqref{szego_ft}.

In the cases of our interest, the emerging singularity is always
a jump discontinuity and, therefore, $\alpha=0$. This value is
not strictly speaking included in the expansion (\ref{asymptotics_sigma_t_small})
of $\zeta(x)$ for $x\to 0$ and,  to our knowledge, the expansion for
$\alpha=0$ has not yet been derived in the literature. We will obtain
it in the following section by employing the general solution of the Painlev\'e V
equation found in~\cite{Gamayun0, Gamayun, Lisovyy} in terms of irregular conformal
blocks \cite{Nagoya}.

\section{Expansion of the Painlev\'e V $\tau$ function}\label{sec:expansion}

In general, the Jimbo-Miwa-Okamoto form of
the Painlev\'e V equation reads \cite{Jimbo}
\begin{multline}\label{painleve_v}
  (x \zeta'')^2=(\zeta-x\zeta'+2(\zeta')^2-2(2\theta_0-\theta_*)\zeta')^2\\
  -4\zeta'(\zeta'-2\theta_0)\left(\zeta'-\theta_0-\theta_t+\theta_*\right)
  \left(\zeta'-\theta_0+\theta_t+\theta_*\right).
\end{multline}
The particular Painlev\'e V equation (\ref{pV_its}), which is relevant
in the study of a Toeplitz determinant with an emergent
singularity, can be obtained choosing in (\ref{painleve_v}),
see also~\cite{Claeys},
\begin{equation}\label{theta_ab_relation}
  \theta_0=0,\quad \theta_t=-\beta,\quad \theta_*=\alpha.
\end{equation}
Hereafter, since we are interested in the limit $\alpha\to 0$, we
will consider $\alpha>0$ without loss of generality.  In principle, there exist other choices for 
$\theta_0$, $\theta_t$ and  $\theta_*$ that produce 
(\ref{pV_its}) but, as we will see, they are  inconsistent with 
(\ref{asymptotics_sigma_t_small}). Finally, all the results that we are going to present are symmetric in the exchange $\beta\rightarrow-\beta$

In order to obtain the solution of (\ref{painleve_v}), 
it is useful to introduce a $\tau$ function that we define as
\begin{equation}\label{tau_function}
  \zeta(x)=x\frac{{\rm d}}{{\rm d}x}\log \tau(x)+(\theta_0-\theta_*)x
  +\theta_0^2-\theta_t^2-2\theta_0\theta_*.
\end{equation}
A combinatorial expansion for $\tau(x)$ was conjectured in \cite{Gamayun0, Gamayun} extending the work \cite{Jimbo}
and eventually proved in \cite{Lisovyy}. 
For further clarifications 
on the notations used, see appendix~\ref{sec:appendixb}. The proposed 
solution is a series representation that involves irregular 
conformal blocks \cite{Nagoya} of a CFT with 
central charge $c=1$. In short, the Painlev\'e V equation 
(\ref{painleve_v}) is solved by the following expansion of $\tau(x)$ 
around $x=0$ \footnote{In order to write the expansion of $\tau(x)$, 
see eq. (4.14) in \cite{Gamayun},
using the same parameters for the boundary conditions $\sigma$ and $s$ 
as in the Theorem 3.1 of \cite{Jimbo}, we have replaced $\theta_*$
by $-\theta_*$ in the structure constants $\mathcal{C}_{\sigma+n}$.
Indeed, one can see that (\ref{tau_lisovyy}) is related to the expansion (4.14) of \cite{Gamayun}
by multiplying and dividing $\mathcal{C}_{\sigma+n}(\theta_0, \theta_t, -\theta_*)$
by $G(1+\theta_*-\sigma-n)G(1+\theta_*+\sigma+n)$, using the identity (see also \cite{Grassi})
$$\frac{G(1-\theta_*-\sigma-n)G(1-\theta_*+\sigma+n)}
{G(1+\theta_*-\sigma-n)G(1+\theta_*+\sigma+n)}=
\frac{G(1-\theta_*-\sigma)(1-\theta_*+\sigma)}
{G(1+\theta_*-\sigma)G(1+\theta_*+\sigma)}
\left(\frac{\sin\pi(\sigma+\theta_*)}{\sin\pi(\sigma-\theta_*)}\right)^n,$$
and redefining the parameter $s$ by absorbing the ratio of sines into it.
The $n$-independent ratio of $G$ functions can be absorbed into the
constant prefactor.}

\begin{equation}\label{tau_lisovyy}
  \tau(x)={\rm const.}\,\sum_{n\in\mathbb{Z}}
  \mathcal{C}_{\sigma+n}(\theta_0,\theta_t, -\theta_*)
  s^nx^{(\sigma+n)^2}\mathcal{B}_{\sigma+n}(x; \theta_0, \theta_t, \theta_*).
\end{equation}
The structure constants $\mathcal{C}_\sigma(\theta_0, \theta_t, \theta_*)$ 
and the irregular conformal blocks $\mathcal{B}_\sigma(x; \theta_0, \theta_t,\theta_*)$ 
are  given by eqs. (4.13) and (4.15) of~\cite{Gamayun}
respectively. The formulas are fully explicit and  will be not repeated  here. The parameters $\sigma$ and $s$ in~\eqref{tau_lisovyy} depend on
the boundary conditions chosen for solving the differential
equation \cite{Jimbo}.  In order to determine their  value we shall compute the first terms of
the expansion of the $\tau$ function and, after inserting
them in (\ref{tau_function}), compare with the asymptotic
expansion (\ref{asymptotics_sigma_t_small}) of $\zeta(x)$
for small $x$.

The first terms of the expansion (\ref{tau_lisovyy})
were actually formerly obtained in~\cite{Jimbo}, cf. Theorem 3.1
therein; according to it\footnote{The relation between the parameters 
for the Painlev\'e V equation (\ref{painleve_v}) chosen in~\cite{Jimbo} 
and the ones used here~\cite{Gamayun} is given in appendix~\ref{sec:appendixb}.}, 
\begin{multline}\label{tau_expansion_jimbo}
  \tau(x)={\rm const.}\,x^{\sigma^2}
  \left[1+\frac{\theta_*(\theta_t^2-\theta_0^2+\sigma^2)}{2\sigma^2}x
  -\hat{s}\frac{(\theta_*+\sigma)((\theta_t-\sigma)^2-\theta_0^2)}
  {4\sigma^2(1+2\sigma)^2}x^{1+2\sigma}\right.
  \\\left.-\hat{s}^{-1}\frac{(\theta_*-\sigma)((\theta_t+\sigma)^2-\theta_0^2)}
  {4\sigma^2(1-2\sigma)^2}x^{1-2\sigma}+O(x^{2-2\sigma})\right],
\end{multline}
where
\begin{equation}\label{hat_s}
  \hat{s}=\frac{\Gamma(1-\theta_*+\sigma)\Gamma(1+\theta_t+\theta_0+\sigma)
  \Gamma(1+\theta_t-\theta_0+\sigma)}{\Gamma(1-\theta_*-\sigma)\Gamma(1+\theta_t+\theta_0-\sigma)
    \Gamma(1+\theta_t-\theta_0-\sigma)}\cdot\frac{\Gamma(1-2\sigma)^2}{\Gamma(1+2\sigma)^2}s,
\end{equation}
and $0<{\rm Re}(2\sigma)<1$.
For the particular set of parameters (\ref{theta_ab_relation})
for the Toeplitz problem, eq. (\ref{asymptotics_sigma_t_small}) 
fixes $\sigma=\theta_*=\alpha$, and the expansion (\ref{tau_expansion_jimbo}) 
simplifies to
\begin{equation}\label{tau_expansion_jimbo_2}
  \tau(x)={\rm const.}\,x^{\alpha^2}
  \left[1+\frac{\alpha^2+\beta^2}{2\alpha}x
  +C_{\alpha,\beta}\frac{\beta^2-\alpha^2}{2\alpha(1+2\alpha)}
  \frac{\sin\pi(\alpha+\beta)}{\sin\pi(\alpha-\beta)}
  sx^{1+2\alpha}+O(x^{2+2\alpha})\right],
\end{equation}
where $C_{\alpha, \beta}$ was defined in (\ref{c_ab}).
Inserting eq. (\ref{tau_expansion_jimbo_2}) 
in (\ref{tau_function}) and comparing the term $O(x^{1+2\alpha})$ 
that is obtained with the corresponding one in the asymptotics (\ref{asymptotics_sigma_t_small}), 
we can conclude that for our problem
$$s=\frac{\sin\pi (\beta-\alpha)}{\sin\pi (\beta+\alpha)}.$$
Once the parameters $\sigma$ and $s$ are fixed, one can 
determine, using the representation (\ref{tau_lisovyy}), the full expansion 
around $x=0$ of the $\tau$ function for the eq. (\ref{pV_its}) in terms of $\alpha$ and $\beta$. 
We have determined  and checked numerically such an expansion up to $O(x^{4+4\alpha})$. Similar reasonings
were done in~\cite{BasorBleher, CarneiroCunha}.

Observe, however, that the expression (\ref{tau_expansion_jimbo_2}) is only valid for 
$2\alpha\not \in \mathbb{Z}$ (i.e. $2\sigma\not\in\mathbb{Z}$).
On the other hand, for the transition between the region $\Sigma_-$ and $\Omega_-$, the symbol 
(\ref{symbol_G}) is never zero, i.e. $\alpha=0$.
We will find now the expansion of the $\tau$ function for this case
studying the limit $\alpha\to0$ in (\ref{tau_expansion_jimbo_2}) and (\ref{tau_lisovyy}). 
The power series in $\alpha$ of the coefficient
$C_{\alpha, \beta}$ around $\alpha=0$ is
\begin{equation}\label{c_expansion}
  C_{\alpha,\beta}=1-2\alpha s_0+(2s_0^2+2-\pi^2/3)\alpha^2+O(\alpha^3),
\end{equation}
where
\begin{equation*}
  s_0=-\psi(1+\beta)-\psi(1-\beta)+3\psi(1)+1
\end{equation*}
with $\psi(z)$ the Digamma function. 
Taking into account (\ref{c_expansion}) and 
$x^{1+n\alpha}\sim (n\alpha\log x+1)x$ for $\alpha\ll 1$, 
from (\ref{tau_expansion_jimbo_2})
and (\ref{tau_lisovyy})
we can conclude that the $\tau$ function 
for the Toeplitz problem~\eqref{pV_its} behaves when $\alpha\to 0$ as
\begin{multline}\label{exp_tau_alpha_0}
  \tau(x)= {\rm const.} 
  \bigg[1-\beta^2x\log x+\beta^2(s_0+1)x
  +\frac{\beta^2}{4}(-3\beta^2+1)x^2  +
  \frac{\beta^4}{12}(\beta^2-1)x^3\log x\\
  -\frac{\beta^2}{72}\Big[\beta^4(3s_0+22)-3\beta^2(s_0+6)+2\Big]x^3
  +\frac{\beta^4(\beta^2-1)}{432}
  \Big[\beta^2(6s_0+17)-6s_0-11\Big]x^4\log x \\
  +\frac{\beta^2}{10368}\Big[36 + \beta^2 
  \big[12 \pi^2 (-1 + \beta^2)^2
  - [655 + 24 s_0 (17 + 6 s_0)] \beta^4
  + 926 \beta^2 +96 s_0 (7 + 3 s_0) \beta^2 \\ 
  -451 - 24 s_0 (11 + 6 s_0) 
  +\big]\Big]x^4
  +O(x^5\log x)\bigg].
\end{multline}

Inserting the expression above in (\ref{tau_function}),
we obtain that for $\alpha=0$
\begin{multline}\label{exp_zeta_alpha_0}
  \zeta(x)= -\beta^2-\beta^2x\log x+s_0\beta^2x
  -\beta^4x^2(\log x)^2+(2s_0+1)\beta^4x^2\log x\\
  +\frac{1}{2}\left(-\beta^4\left(2s_0(s_0+1)+3\right)
  +\beta^2\right)x^2
  +O(x^3(\log x)^3).
\end{multline}

We will employ in the next section the Painlev\'e V $\tau$ function (\ref{exp_tau_alpha_0}) in the analysis of the behaviour of the EFP in the double scaling limit $L\to \infty$ and $|h|\to 1$.

\section{Exact result for $\gamma=1$ and conjectures for $\gamma\neq 1$}\label{sec:results}

In this section, we will apply and generalize 
the mathematical results of the two previous 
sections to the EFP in the XY spin chain for 
$\gamma\neq 0$. We will consider separately the 
transitions discussed in section \ref{sec:transition} from the regions
$\Sigma_-$ and $\Sigma_0$ to the critical lines 
$\Omega_-$ and $\Omega_+$ respectively. In each case, 
the strategy will be first to study the quantum Ising 
chain ($\gamma=1$), where the interpolation formula 
(\ref{ick_asymptotics}) can be directly applied, and 
then to extend it to $\gamma\neq 1$ exploiting the 
properties of the EFP under certain M\"obius transformations 
that act on the parameters of the chain $h$, $\gamma$.

\begin{description}
\item \textit{Transition from $\Sigma_-$ to the critical line $\Omega_-$.}

  For the quantum Ising line in the region $\Sigma_-$, 
  the symbol (\ref{symbol_G}) has exactly the form of the prototype 
  (\ref{ick_symbol}) considered in \cite{Claeys} 
  with $\alpha=0$, $\beta=1/4$, $t=\log|h|$ and $Q(z)=\log(h^{-1/4} V(z))+\ii\pi/4$.
  Thus we can directly apply the asymptotics (\ref{ick_asymptotics}).
  Comparing the expression for $\tilde{\Omega}$ 
  given in eq. (\ref{omega_tilde}) 
  with the definition (\ref{tau_function}) of the Painlev\'e 
  V $\tau$ function, in our case we have 
  $$\tilde{\Omega}(x)=\log\tau(x)$$
  because $\alpha=0$. Notice  that $\tilde{\Omega}(0)=0$,
  therefore the constant in (\ref{exp_tau_alpha_0}) is equal to one.
  If we also take into account that the 
  function $V(\e^{\ii \theta})$ is even and, consequently,
  $V_{-k}=V_{k}$, we find that eq.~(\ref{ick_asymptotics}) leads to
  \begin{multline}\label{scaling_lim_ising_sigma_-}
    \log \mathcal{P}(L, h, 1)=A(h, 1) L
    -\frac{1}{16}\log(2L\log|h|)+\log \tau(2L\log|h|)\\
    +E[V]+\frac{1}{16}\log(1-h^{-2})
    +\log\left[G\left(\frac{5}{4}\right)G\left(\frac{3}{4}\right)\right]
    +o(1),
  \end{multline}
  for $h<-1$.

  Substituting then eq.~(\ref{exp_tau_alpha_0}) into~\eqref{scaling_lim_ising_sigma_-} and calling $x=2L\log |h|$, we obtain finally 
  \begin{multline}\label{tau_small}
 \log\tau(x)\stackrel{x\ll 1}{\sim} 
 -\frac{1}{16}x\log x
 +\frac{\tilde{\gamma}_{\rm E}-1}{8}x
 -\frac{1}{512}x^2(\log x)^2
 +\frac{\tilde{\gamma}_{\rm E}-1}{128}x^2\log x
 +\frac{1}{64}\left(-\frac{1}{2}\tilde{\gamma}_{\rm E}^2+\tilde{\gamma}_{\rm E}+\frac{5}{16}\right)x^2\\
 - \frac{1}{12288}x^3 (\log x)^3
  + \frac{1}{2048} (\tilde{\gamma}_{\rm E} - 1)x^3 (\log x)^2 
 - \frac{1}{1024}\left(\tilde{\gamma}_{\rm E}^2 - 2\tilde{\gamma}_{\rm E} + \frac{1}{2}\right)x^3\log x  \\
 + \left(\frac{\tilde{\gamma}_{\rm E}^3}{1536} - \frac{\tilde{\gamma}_{\rm E}^2}{512} 
    + \frac{11\tilde{\gamma}_{\rm E}}{16384} - \frac{35}{98304}\right)x^3 
 - \frac{1}{262144}x^4(\log x)^4
 + \frac{\tilde{\gamma}_{\rm E} - 1}{32768}x^4(\log x)^3 \\
 - \left(\frac{3\tilde{\gamma}_{\rm E}^2}{32768} 
    - \frac{3\tilde{\gamma}_{\rm E}}{16384} + \frac{1}{16384}\right)x^4 (\log x)^2 
 + \left(\frac{\tilde{\gamma}_{\rm E}^3}{8192} - \frac{3\tilde{\gamma}_{\rm E}^2}{8192}+ 
    \frac{21\tilde{\gamma}_{\rm E}}{65536} - \frac{319}{3145728}\right)x^4\log x \\
 +\left(-\frac{\tilde{\gamma}_{\rm E}^4}{16384} 
   +\frac{\tilde{\gamma}_{\rm E}^3}{4096} 
   - \frac{13\tilde{\gamma}_{\rm E}^2}{32768} 
   + \frac{709\tilde{\gamma}_{\rm E}}{1572864} 
   + \frac{25\pi^2}{6291456} -\frac{16099}{75497472}\right)x^4 
\end{multline}
where $\tilde{\gamma}_{\rm E}=3\log 2-\gamma_{\rm E}/2$ and $\gamma_{\rm E}$ is
the Euler-Mascheroni constant.
  
  On the other hand, for $L\log|h|\to\infty$ we should have
  \begin{equation}\label{tau_large}
  \log \tau(x)\stackrel{x\gg 1}{\sim} \frac{1}{16}\log(x)
  -\log\left[G\left(\frac{5}{4}\right)G\left(\frac{3}{4}\right)\right]
  \end{equation}
  in order to recover the asymptotics (\ref{asymp_sigma_-})
  predicted by the Szeg\H{o} theorem in the region $\Sigma_-$.
  
  For the theories in the region $\Sigma_-$ outside the 
  quantum Ising line, $\gamma=1$, the symbol $\mathcal{G}(z)$ 
  is not of the form (\ref{ick_symbol}) and, therefore, 
  we cannot directly apply the interpolation formula
  (\ref{ick_asymptotics}). Nevertheless, we are now going 
  to conjecture a generalization of eq. (\ref{scaling_lim_ising_sigma_-}) 
  for $\gamma\neq 1$ that we will check numerically later.
  First, let us relate the symbol $\mathcal{G}(z)$ to one of the 
  form (\ref{ick_symbol}) through a M\"obius transformation.
  Consider the subgroup of these transformations of the form
  \begin{equation}\label{moebius}
    z'=\frac{z\cosh \delta +\sinh \delta}
    {z\sinh\delta+\cosh\delta}, \quad 
    \delta\in \mathbb{R},
  \end{equation}
  that preserve the real line and the unit circle 
  and map the unit disk and the upper half plane 
  into themselves. Now, if we consider a chain with 
  parameters $h<-1$ and $\gamma\neq 1$, then the 
  branch points of the corresponding symbol 
  $\mathcal{G}(z)$ are the points on the real line 
  $(z_+, z_-, z_-^{-1}, z_+^{-1})$ with $z_\pm$ 
  defined in (\ref{branch_points}). For $\tanh\delta=-z_+$, 
  the transformation (\ref{moebius}) maps these branch points
  to the set of points $(0, h', 1/h', \infty)$, with $h'$ 
  such that
  \begin{equation}\label{inv_moebius}
    h'^2-1=\frac{h^2-1}{\gamma^2}.
  \end{equation}
  Under the transformation (\ref{moebius}), the Laurent polynomials 
  (\ref{laurent}) change to \cite{Vilenkin}
  $$\Phi'(z)=\frac{z}{2}-h'+\frac{z^{-1}}{2},\quad
  \Xi'(z)=\frac{1}{2}(z-z^{-1}).$$
  These new Laurent polynomials define the transformed symbol
  $\mathcal{G}'$ associated to the quantum Ising chain with 
  transverse magnetic field $h'$. The new symbol is related
  to the old one with parameters $h$, $\gamma$, i.e. $\mathcal{G}(z)$,
  as
  \begin{equation}\label{moebius_G}
    \mathcal{G'}(z')=\mathcal{G}(z).
  \end{equation}
  
  In \cite{Ares1, Ares2}, the authors studied the behaviour under
  the transformations (\ref{moebius}) of the R\'enyi entanglement
  entropy in the ground state of the XY spin chain. This entropy
  can be written in terms of a block Toeplitz determinant with a 
  symbol also obeying the identity (\ref{moebius_G}). Inspired by 
  the results found there, we establish the following conjecture for 
  the behaviour of the EFP under (\ref{moebius}). Let us substract 
  from $\mathcal{P}$ the contribution of the factor $V(\ee^{\ii\theta})$ which 
  appears in the symbol $g(\theta)$; namely define
   \begin{equation}\label{def_e-}
    \mathcal{E}^-(x, h, \gamma)\equiv{\rm e}^{-A(h, \gamma) L-E[V]}
    \mathcal{P}(L, h, \gamma).
   \end{equation}
  Then we conjecture that, for large $L$, 
  \begin{equation}\label{moebius_e_-}
  \mathcal{E}^-(x, h, \gamma)=\mathcal{E}^-(x/\gamma, h', 1);
  \end{equation}
  that is, we assume that the quantity $\mathcal{E}^-$ transforms 
  under (\ref{moebius}) in the same way as the entanglement
  entropy of the ground state of the chain, see in particular
  eqs. (45) and (46) of~\cite{Ares2}.
  Inserting (\ref{scaling_lim_ising_sigma_-}) in the
  right hand side of (\ref{moebius_e_-}) and using 
  (\ref{inv_moebius}), we conclude
  \begin{multline}\label{e_-}
  \log\mathcal{E}^-(x, h, \gamma)=-\frac{1}{16}\log\frac{x}{\gamma}
  +\log \tau\left(\frac{x}{\gamma}\right)\\
  -\frac{1}{16}\log\left(1-\frac{\gamma^2}{1-h^2}\right)
  +\log\left[G\left(\frac{3}{4}\right)G\left(\frac{5}{4}\right)\right]+o(1).
  \end{multline}
  Physically, the transformation (\ref{moebius_e_-}) 
  implies that, for any value of the anisotropy parameter 
  $\gamma\neq 0$, the double scaling limit $h\to -1^-$, $L\to\infty$
  is described by the same Painlev\'e V $\tau$ function as in the
  quantum Ising line. This is very reasonable \cite{Tracy-review} since all the points 
  in the line $h=-1$ belong to the Ising criticality class. Observe
  that, taking into account (\ref{tau_large}), the rescaling by $\gamma^{-1}$ 
  of the argument of the $\tau$ function in (\ref{e_-}) is the simplest way of 
  cancelling the logarithmic term $-1/16\log(x/\gamma)$
  in the limit $x\to \infty$ and recovering the prediction 
  of the Szeg\H{o} theorem (\ref{asymp_sigma_-}).  
  Note also that, in virtue of (\ref{inv_moebius}), the terms in the second 
  line of the expression (\ref{e_-}) are invariant under 
  the transformation (\ref{moebius}).  
  
  Finally, plugging (\ref{e_-}) into (\ref{def_e-}),
  we can conclude that
  \begin{multline}\label{interpolation_efp_sigma_-}
    \log \mathcal{P}(L, h, \gamma)=A(h,\gamma) L-\frac{1}{16}\log\frac{2L\log|h|}{\gamma}
    +\log \tau\left(\frac{2L\log|h|}{\gamma}\right)\\
    +E[V]-\frac{1}{16}\log\left(1-\frac{\gamma^2}{1-h^2}\right)
    +\log\left[G\left(\frac{5}{4}\right)G\left(\frac{3}{4}\right)\right]+o(1). 
  \end{multline}
  In fig. \ref{fig:efp_xy_sigma_-}, we study numerically the interpolation between
  the asymptotic behavior of $\log \mathcal{P}(L, h, \gamma)$ at $h=-1$
  and for $h<-1$. In particular, we consider the quantity
  \begin{equation}\label{delta_-}
  \Delta^-(x, h,\gamma)=\log \mathcal{E}^-(x, h, \gamma)+\frac{1}{16}\log\frac{2L\log|h|}{\gamma}
  +\frac{1}{16}\log\left(1-\frac{\gamma^2}{1-h^2}\right)
    -\log\left[G\left(\frac{5}{4}\right)G\left(\frac{3}{4}\right)\right],
  \end{equation}
  as a function of $2L\log |h|/\gamma$ for several values of $L$ and $\gamma$.
  Replacing in the expression above $\log\mathcal{E}^-$ by (\ref{e_-}), we expect that
  $$\Delta^-(x, h,\gamma)=\log \tau\left(\frac{2L\log|h|}{\gamma}\right)$$
  in the double scaling limit $h\to -1^-$ and $L\to \infty$.
   \begin{figure}[t]
  \centering
  \includegraphics[width=0.8\textwidth]{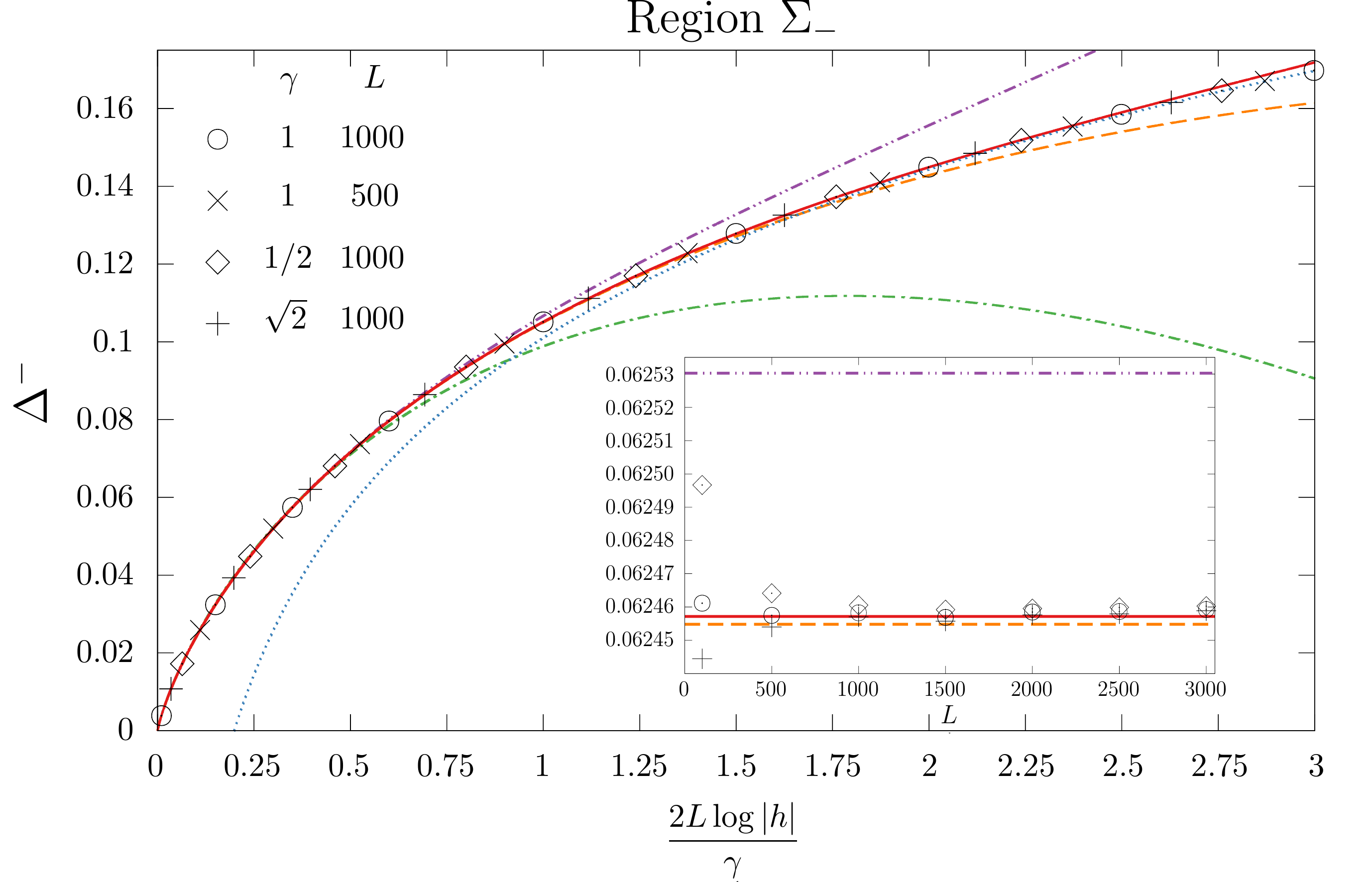}
     \caption{Difference $\Delta^-$ between the logarithm 
    of the EFP for $h<-1$ (region $\Sigma_-$) and all the terms 
    in the interpolation formula (\ref{interpolation_efp_sigma_-}) 
    except the one containing the $\tau$ function, see (\ref{delta_-}), as 
    a function of the ratio $x/\gamma$, with $x=2L\log|h|$. It is expected 
    that $\Delta^-=\log\tau(x/\gamma)$ for large $L$. The dots
    have been obtained computing numerically the EFP for several values 
    of $L$ and $\gamma$ and varying the magnetic field $h$. The dot-dashed lines 
    $\textcolor{mygreen}{\mydashedlinea}$, $\textcolor{myviolet}{\mydashedlinec}$, and 
    $\textcolor{myorange}{\mydashedlined}$ are the expansion (\ref{tau_small}) 
    of $\log \tau(x/\gamma)$ for small $x/\gamma$ considering
    only the terms up to order $O(x/\gamma)$, $O(x^2/\gamma^2)$ and
    $O(x^3/\gamma^3)$ respectively. 
    The solid line $\textcolor{myred2}{\mysolidline}$ corresponds to 
    the same expansion but taking all the terms found in (\ref{tau_small}). 
    The dotted line $\textcolor{myblue}{\mydashedlineb}$ is the behaviour (\ref{tau_large}) 
    of $\log \tau(x/\gamma)$ for large $x/\gamma$
    In the inset, we have fixed the value of the scaling variable $x/\gamma=0.4$ and 
    we have represented $\Delta_-$ as a function of the length $L$ for
    $\gamma=1\,(\mycircle), 1/2\,(\mydiamond)$ and $\sqrt{2}\,(+)$. In this case, the straight lines
    indicate the value of $\log\tau(0.4)$ computed using (\ref{tau_small}),
    taking the terms up to second ($\textcolor{myviolet}{\mydashedlinec}$), third
    ($\textcolor{myorange}{\mydashedlined}$), and fourth ($\textcolor{myred2}{\mysolidline}$) order.}
    
  \label{fig:efp_xy_sigma_-}
   \end{figure}
  The points in fig. \ref{fig:efp_xy_sigma_-} have been obtained computing numerically
  $\mathcal{P}(L, h, \gamma)$ through the determinant (\ref{efp_det}).
  This determinant has been calculated from the eigenvalues of the matrix
  $S$ whose entries are given by (\ref{matrix_s}) in the thermodynamic
  limit. Then we have applied the definition (\ref{def_e-}) to determine $\mathcal{E}^-(x, h, \gamma)$.
  The coefficients $A(h, \gamma)$ and $E[V]$ have been obtained computing
  by numerical integration the Fourier modes of $V(z)$. The sum in $E[V]$
  has been evaluated up to $10^4$ modes. The curves in fig. 
  \ref{fig:efp_xy_sigma_-} are the expansions (\ref{tau_small}) and (\ref{tau_large}) 
  of $\log\tau(2L\log h/\gamma)$ for small and large $2L\log|h|/\gamma$.
  There is an excellent agreement between the numerical points and the 
  expansion of $\log\tau(2L\log|h|/\gamma)$ obtained in (\ref{tau_small}).
  Observe that the points in the figure only depend on the product 
  $2L\log|h|/\gamma$. This fact strongly supports the transformation 
  (\ref{moebius_e_-}) proposed for $\mathcal{E}^-$ under the M\"obius 
  transformations (\ref{moebius}). In order to explicitly check the universality
  of the $\tau$ function, in the inset of fig.~\ref{fig:efp_xy_sigma_-} we have taken 
  a fixed value for the scaling variable $x/\gamma$ and calculated 
  $\Delta_-$ in terms of the length $L$ for different anisotropies $\gamma$. If the 
  conjecture (\ref{moebius_e_-}) is correct, then the points corresponding to different
  $\gamma$ should collapse to the same value of $\Delta_-$ at $L\to\infty$.
  As is clear in the inset, the distance between the points of different
  $\gamma$ shrinks as $L$ increases. Notice that for $L=3000$ the 
  relative differences between them are smaller than $10^{-5}$.

\item \textit{Transition from $\Sigma_0$ to the critical line $\Omega_+$}
  
  Recall that in the region $\Sigma_0$ the symbol (\ref{g_sigma_0})
  presents a Fisher-Hartwig singularity at $\theta=\pi$;
  for the quantum Ising line, $\gamma=1$, reduces to
  $$g(\theta)=W^0(\theta)\left(\frac{1-h\e^{-\ii\theta}}
  {1-h\e^{\ii\theta}}\right)^{1/4}
  (2-2\cos(\theta-\pi))^{1/2}
  \e^{\ii/2(\theta-\pi-\pi\sign(\theta-\pi))}.$$
  Except for the two last factors,
  that correspond to the zero and the discontinuity
  at $\theta=\pi$ respectively, the symbol is of the form
  (\ref{ick_symbol}) studied in \cite{Claeys} with $\alpha=0$, $\beta=1/4$, $t=-\log h$
  and $Q(\e^{\ii\theta})=\log(h^{-1/4} W^0(\theta))+\ii \pi/4$.
  According to the localization theorem \cite{Basor1}, see appendix \ref{sec:appendixa}, in the double scaling limit
  $h\to 1^-$, $L\to \infty$, the Fisher-Hartwig
  singularity at $\theta=\pi$ can be ignored and the interpolation
  formula (\ref{ick_asymptotics}) continues to hold.
  From the physical point of view (compare figs. \ref{fig:scaling_limit_sigma_-} and
  \ref{fig:scaling_limit_sigma_0}), this is again natural if we require that a unique $\tau$ function 
  should describe the crossover of the EFP between the off-critical 
  and critical regimes in the Ising universality class.
  Therefore, we should have
  \begin{multline}\label{scaling_lim_ising_sigma_0}
    \log \mathcal{P}(L, h, 1)=
    A(h, 1)L
    -\frac{1}{16}\log\left(-2L\log h \right)
    +\log\tau(-2L\log h)\\+E[W^0]
    -W_-^0(\pi)
    +\frac{1}{16}\log\frac{1-h^2}{(1+h)^4}
    +\log\left[G\left(\frac{3}{4}\right)G\left(\frac{5}{4}\right)\right]
    +o(1).
  \end{multline}

In order to obtain the expansion of $\mathcal{P}$
for $\gamma\neq 1$ in the double scaling limit, we again
exploit the M\"obius symmetry as in the transition
from $\Sigma_-$ to $\Omega_-$. Considering the expansion
(\ref{asymp_sigma_0}) given by the Fisher-Hartwig
conjecture in the region $\Sigma_0$, let us define now
the quantity
\begin{equation}\label{def_e0}
  \mathcal{E}^0(x, h, \gamma)\equiv
  \e^{-A(h, \gamma)L-E[W^0]+W_-^0(\pi)+\frac{1}{4}\log\frac{1+h}{\gamma}}
  \mathcal{P}(L, h, \gamma).
\end{equation}
Namely, we remove from the EFP the contribution of the factor $W^0(\theta)$ in the symbol as well as
the one of
the Fisher-Hartwig singularity at $\theta=\pi$.
Then, analogously to what we have done in (\ref{moebius_e_-})
for the region $\Sigma_-$, we conjecture that
\begin{equation}\label{moebius_e_0}
\mathcal{E}^0(x, h, \gamma)=
\mathcal{E}^0(x/\gamma, h', 1),
\end{equation}
for $L\to\infty$.
 
Combining this conjecture with (\ref{scaling_lim_ising_sigma_0})
and taking into account the identity (\ref{inv_moebius}) between 
$h$, $\gamma$ and $h'$, we conclude that
\begin{multline}\label{e_0}
  \log\mathcal{E}^0(x, h, \gamma)=
  -\frac{1}{16}\log\left(-\frac{x}{\gamma}\right)
  +\log\tau\left(-\frac{x}{\gamma}\right)\\
  +\frac{1}{16}\log\frac{1-h^2}{\gamma^2}
  +\log\left[G\left(\frac{3}{4}\right)G\left(\frac{5}{4}\right)\right]
  +o(1).
\end{multline}
Finally, plugging this result into (\ref{def_e0}), we arrive at
\begin{multline}\label{interpolation_efp_sigma_0}
  \log\mathcal{P}(L, h, \gamma)=A(h, \gamma)L
  -\frac{1}{16}\log\left(-\frac{2L\log h}{\gamma}\right)+
  \log\tau\left(-\frac{2L\log h}{\gamma}\right)
  \\+\frac{1}{16}\log\frac{(1-h^2)\gamma^2}{(1+h)^4}
  +E[W^0]-W_-^0(\pi)+
  \log\left[G\left(\frac{3}{4}\right)G\left(\frac{5}{4}\right)\right]+o(1).
\end{multline}

 \begin{figure}[t]
  \centering
  \includegraphics[width=0.8\textwidth]{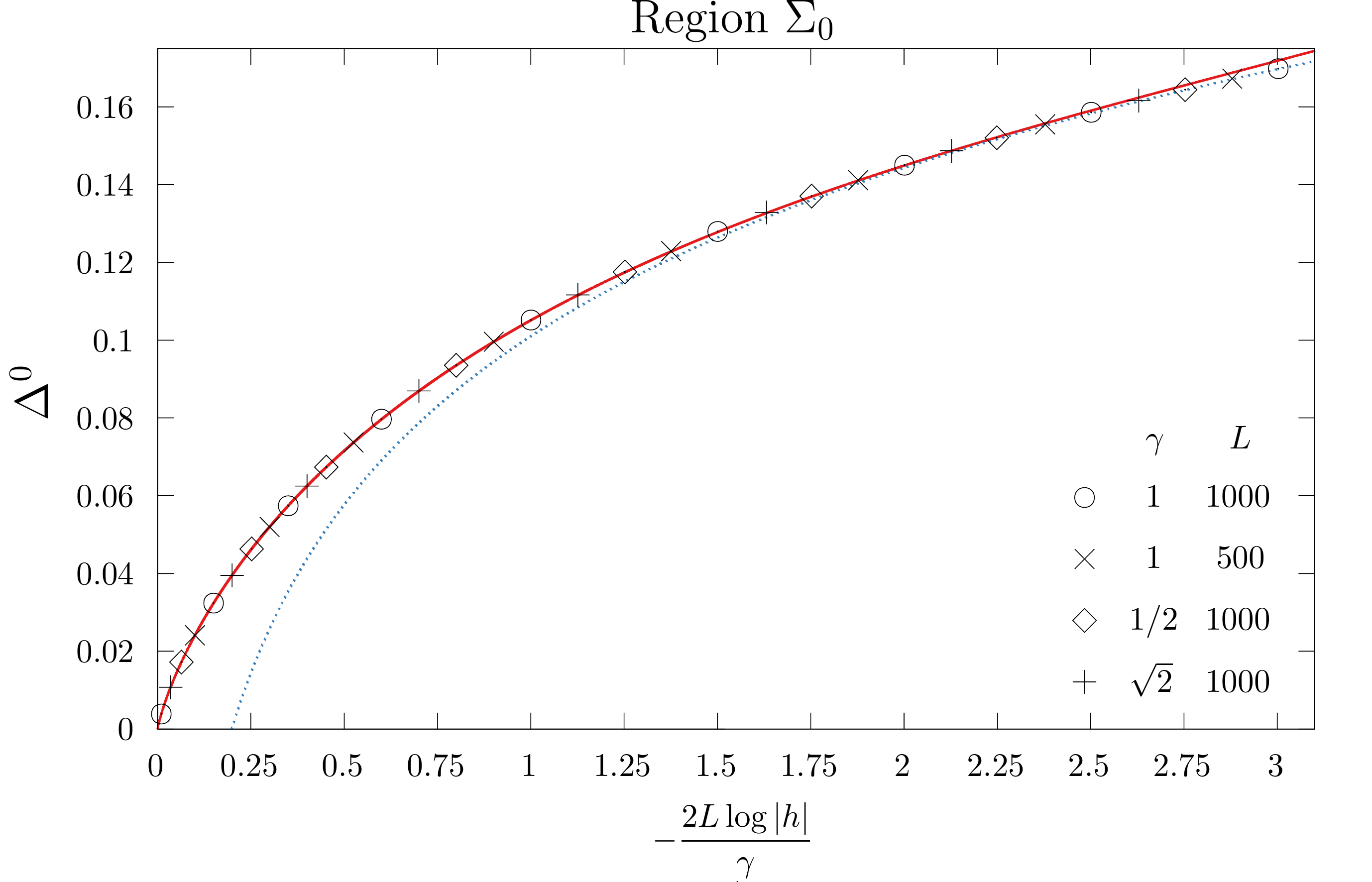}
     \caption{We represent the difference $\Delta^0$ between
    the logarithm of the EFP for $|h|<1$ (region $\Sigma_0$)
    and the terms in the interpolation conjectured in 
    (\ref{interpolation_efp_sigma_0}) except the one that
    contains the Painlev\'e V $\tau$ function, see (\ref{delta_0}).
    Thus $\Delta^0$ should be asymptotically equal to 
    $\log\tau(-x/\gamma)$ where $x=2L\log|h|$. The points 
    were found calculating numerically the EFP for different 
    $L$ and anisotropies $\gamma$. The solid line 
    $\textcolor{myred2}{\mysolidline}$ is the expansion 
    (\ref{tau_small}) of $\log\tau(-x/\gamma)$ around 
    $x/\gamma=0$. 
    The dotted curve $\textcolor{myblue}{\mydashedlineb}$ is 
    the asymptotic behaviour (\ref{tau_large}) of $\log\tau(-x/\gamma)$ 
    for large $-x/\gamma$.
    }
  \label{fig:efp_xy_sigma_0}
   \end{figure}

In fig. \ref{fig:efp_xy_sigma_0} we analyze numerically 
the transition from the non-critical region $\Sigma_0$ 
to the critical line $\Omega_+$. Analogously to the previous 
section, we consider the quantity
\begin{equation}\label{delta_0}
\Delta^0(x, h, \gamma)=\log\mathcal{E}^0(x, h, \gamma)
+\frac{1}{16}\log\left(-\frac{x}{\gamma}\right)
  -\frac{1}{16}\log\frac{1-h^2}{\gamma^2}
  -\log\left[G\left(\frac{3}{4}\right)G\left(\frac{5}{4}\right)\right]
\end{equation}
and plot it as a function of $-2L\log h/\gamma$.
According to (\ref{e_0}), we expect 
$$\Delta^0(x, h, \gamma)= \log\tau\left(-\frac{2L\log h}
{\gamma}\right)$$
in the limit $L\to\infty$.

The points in fig. $\ref{fig:efp_xy_sigma_0}$ have been obtained
computing numerically $\mathcal{P}(L)$, $A(h, \gamma)$ and the 
series in $E[W^0]$ and $W_-^0(\pi)$ as described in the case of the
region $\Sigma_-$. Here the series have been evaluated up to $10^4$ modes. 
The curves correspond to the asymptotics of the logarithm of 
the $\tau$ function that we have found in (\ref{tau_small}) and 
(\ref{tau_large}) for small and large $-2L\log |h|/\gamma$. 
There is an excellent agreement between the numerical points and the 
asymptotic expansions. The numerics provides again an unbiased and
strong support to the interpolation formula (\ref{interpolation_efp_sigma_0}). 
Note that this is also a test of the transformation (\ref{moebius_e_0}) 
of $\mathcal{E}^0$ under the M\"obius transformations (\ref{moebius}).

\end{description}

\section{The XX spin chain}\label{sec:xx}

\begin{figure}
  \centering
  \includegraphics[width=0.8\textwidth]{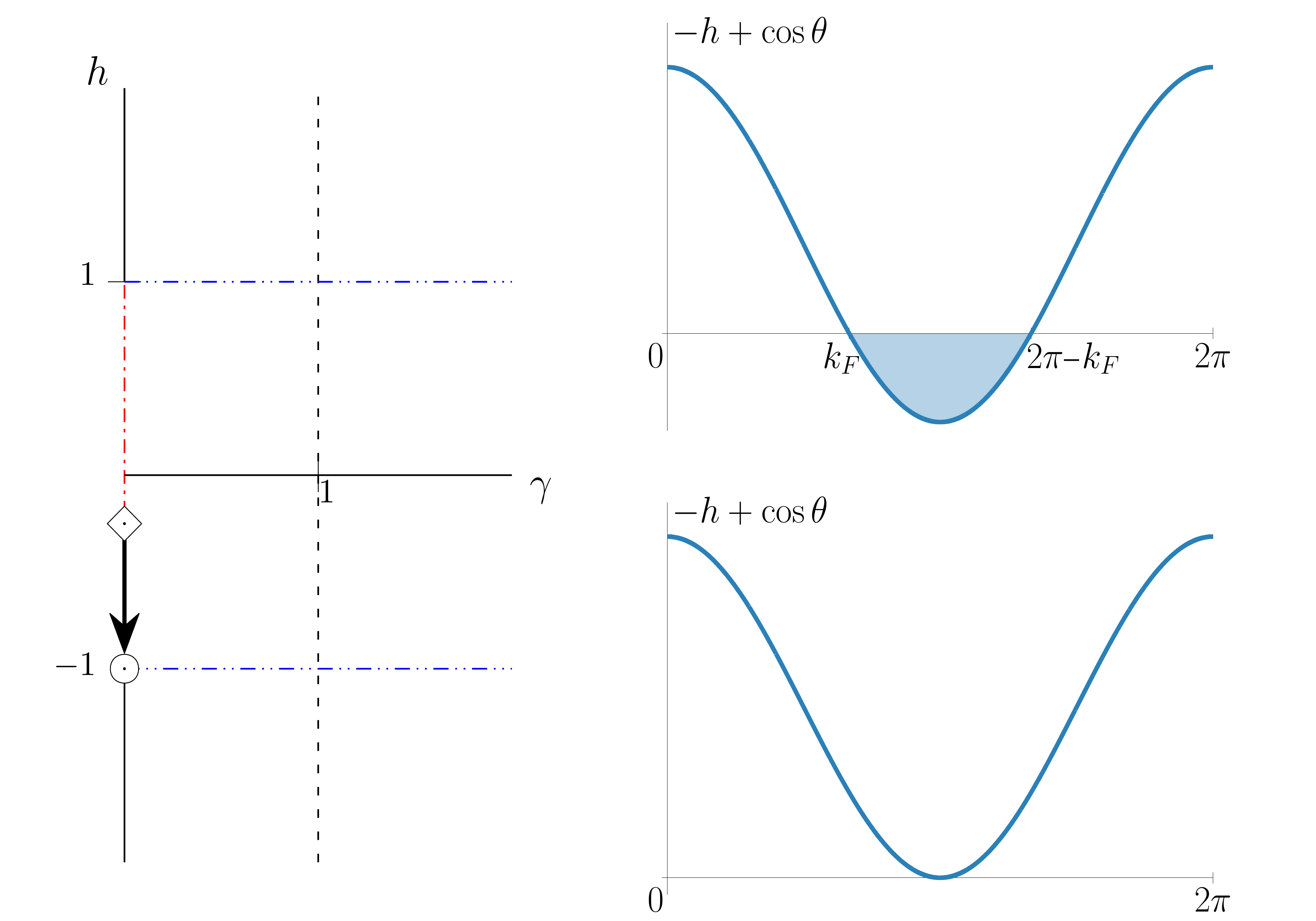}
  \caption{In a XX spin chain with $|h|<1$
    (e.g. the point $\mydiamond$ in the plane
    $(\gamma, h)$), the mass gap is zero and the dispersion relation
    is like that represented in the upper right panel: it is negative
    in the interval $(k_F, 2\pi-k_F)$.
    As $h\to -1^+$, the dispersion relation moves vertically upwards,
    the Fermi momentum $k_F\to0$, and the number of particles in the Dirac
    sea decreases. At $h=-1$, the point $\mycircle$, the
    dispersion relation is that plotted in the lower right panel.
    It becomes non-negative and the Dirac sea disappears.
    }
   \label{fig:scaling_limit_omega_0}
   \end{figure}

As we have already discussed in sections~\ref{sec:intro} and \ref{sec:efp}, and it was firstly found
in \cite{Shiroishi}, the EFP behaves differently when
$\gamma$ vanishes (XX spin chain) and the magnetization is conserved. 
For $|h|<1$, the chain is critical, and the ground state is a Dirac sea, in which $\mathcal{P}_0(L, k_F)$ displays a Gaussian
decay with $L$, see (\ref{efp_omega_0}). On the other hand, for $|h|>1$, there is
no Dirac sea, and the EFP is either 0 or 1 for $h>1$
and $h<-1$ respectively. The limit $|h|\to 1^-$ is both mathematically
and physically different from the ones studied previously when
$\gamma\neq 0$. Mathematically, for $|h|<1$, the symbol of $S$ is the piecewise
constant function (\ref{symbol_xx}) with support $[k_F, 2\pi- k_F]$, where
$k_F$ is the Fermi momentum, $k_F=\arccos h$. From the physical side, in the limit $|h|\to 1^-$ the fermion
density $\frac{k_F}{\pi}\to 0$ and the ground state is no longer the
Dirac sea but the trivial Fock vacuum, see fig. \ref{fig:scaling_limit_omega_0}. 

In order to study the double scaling limit $L\to \infty$, $|h|\to 1^-$,
let us first observe that for $|h|<1$, the entries of the
Toeplitz matrix $S$ are
\begin{equation}\label{matrix_s_omega_0}
  S_{nm}=\frac{\sin[k_F(n-m)]}{\pi(n-m)}, \quad S_{nn}=\frac{k_F}{\pi}.
\end{equation}
Reintroducing the lattice spacing $a$, we can parametrize
$L=\ell a^{-1}$, $n=ya^{-1}$ and $k_F=p_Fa$; therefore, the double scaling 
limit $L\to\infty$, $k_F\to 0$ is nothing but the continuum 
limit  $a\to0$ with $\chi=Lk_F$ fixed.  
Then, following refs. \cite{Dyson, Krasovsky, Ivanov, Abanov2},
in the $a\to 0$ limit, the EFP turns out to be the Fredholm determinant 
\begin{equation}\label{fredholm}
  \mathcal{P}_0(L, k_F)=|\det(I-K_{\rm sine}^\chi)|,
\end{equation}
where $K_{\rm sine}^\chi$ is the trace-class operator acting on $L^2([0,\chi])$
with kernel
$$K_{\rm sine}(y, y')=\frac{\sin(y-y')}{\pi(y-y')}.$$
This is the sine kernel \cite{Dyson}, well known in the theory of random matrices~\cite{Metha_book}. In the literature are already present several detailed analysis of the Fredholm determinant (\ref{fredholm}). 
For instance, one could set up an expansion similar to \cite{Perk1, Perk2, Perk3}
for the time-dependent correlation functions of the critical Ising
chain or to \cite{Ivanov} for
the full-counting statistics by solving a Riemann-Hilbert problem.
On the other hand, in the celebrated paper \cite{Jimbo2}, Jimbo,
Miwa, Mori and Sato discovered that the Fredholm determinant (\ref{fredholm}) is equal to the $\tau$ function, $\tau_0$,
of the Painlev\'e V equation (\ref{painleve_v}) with
$x=\ii2\chi$, parameters 
$$\theta_0=\theta_t=\theta_*=0,$$
and satisfying the boundary conditions
\begin{equation}\label{asymptotics_zeta_0}
\zeta_0(\chi)\sim -\frac{\chi}{\pi}-\frac{\chi^2}{\pi^2},
\quad \mbox{for}\quad \chi\to 0.
\end{equation}
Therefore, we can conclude that 
\begin{equation}\label{efp_painleve_omega_0}
  \mathcal{P}_0(L, k_F)=\tau_0(L k_F),
\end{equation}
in the double scaling limit $L\to\infty$ and $k_F\to 0$.
The Painlev\'e V $\tau$ function $\tau_0(Lk_F)$ describes
the crossover of the EFP from a theory whose low energy fermionic excitations have a linear dispersion to another theory with quadratic dispersion, see fig. \ref{fig:scaling_limit_omega_0}.

The expansion of $\tau_0(\chi)$ around $\chi=0$
can be computed from the general expression (\ref{tau_lisovyy})
for the Painlev\'e V $\tau$ function by  a suitable limiting procedure. 
Let consider the  Painlev\'e
V equation (\ref{painleve_v}) with parameters
$$\theta_t=-\frac{\sigma}{2},\quad
\theta_0=\frac{\sigma}{2},\quad \mbox{and} \quad 
\theta_*\neq 0,$$
and then take the limit $\sigma\to 0$. 

With this choice of the parameters in (\ref{tau_lisovyy}), we obtain in the limit $\sigma\rightarrow 0$
\begin{multline}
\label{tau_int}
 \tau(x)= {\rm const.}\left[1+\frac{(-s+1)\theta_*}{2}x
+\frac{(-s+1)\theta_*^2}{4}x^2+\frac{(-s+1)\theta_*^3}{12}x^3\right.\\
\left.+\frac{(-s^2+1+\theta_*^2(s^2-12s+11))\theta_*^2}{576}x^4
+O(x^5)\right].
\end{multline}
The expression above for the $\tau$ function leads to the asymptotics (\ref{asymptotics_zeta_0}) by identifying
\begin{equation}
\label{sine_s}
s=-\frac{\ii}{\pi\theta_*},
\end{equation}
and  taking the limit $\theta_*\to 0$. Indeed, substituting $x=\ii 2\chi$ and~\eqref{sine_s} into \eqref{tau_int}  we obtain
\begin{equation}
\label{tau_sine}
\tau_0(\chi)={\rm const.}\left(1-\frac{\chi}{\pi}
+\frac{\chi^4}{36\pi^2}+O(\chi^6)\right),
\end{equation}
which, plugged into (\ref{tau_function}) gives back (\ref{asymptotics_zeta_0}).
This expansion coincides with that obtained in \cite{Forrester}, cf. eq. (8.114)
upon replacing $\xi=1$ and $t=\chi/\pi$. 
Equivalently~\cite{Gamayun}, eq.~\eqref{tau_sine} could be also derived  directly from the combinatorial expansion for the Painlev\'e VI $\tau$ function \cite{Gamayun0} after a series of confluent limits.

\begin{figure}[t]
  \centering
  \includegraphics[width=0.8\textwidth]{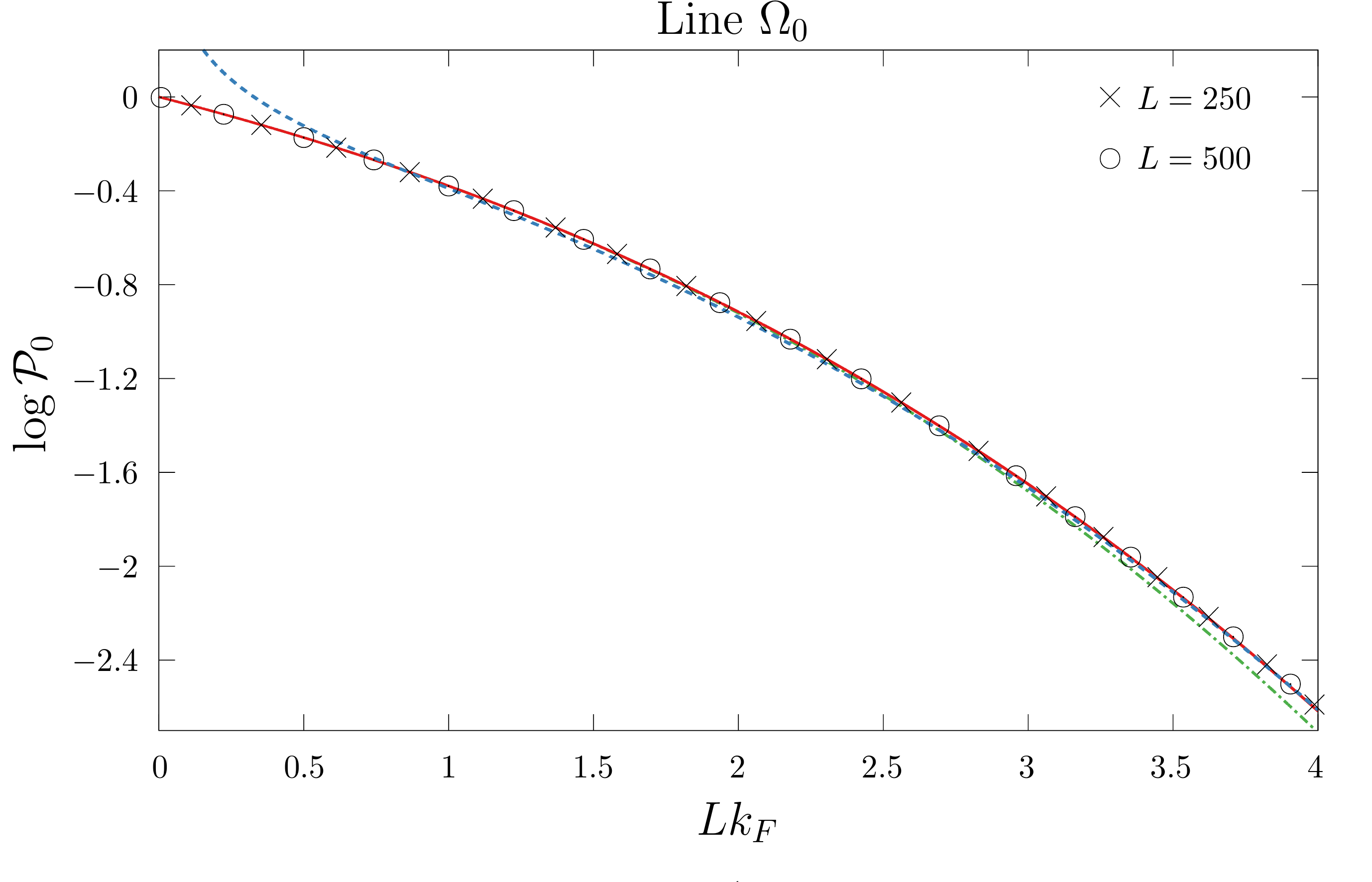}
  \caption{Logarithm of the EFP
  for the ground state of the critical XX spin chain (line 
  $\Omega_0$: $\gamma=0$, $|h|<1$) as a function of the product 
  $Lk_F$. The dots have been obtained numerically 
  taking two fixed values of $L$ and varying $k_F$. It is expected 
  that, in the double scaling limit $L\to\infty$, $k_F\to0$, 
  $\mathcal{P}_0(L, k_F)$ is described by 
  the Painlev\'e V $\tau$ function $\tau_0(Lk_F)$, see eq. 
  (\ref{efp_painleve_omega_0}). The dotted-dashed line 
  $\textcolor{mygreen}{\mydashedlinea}$ represents the expansion 
  (\ref{tau0_small}) of $\log \tau_0(Lk_F)$ around $Lk_F=0$. The solid 
  line $\textcolor{myred2}{\mysolidline}$ corresponds to the same expansion (\ref{tau0_small}) but adding the 
  terms up to $O((Lk_F)^6)$ that can be determined from  
  eq. (8.114) of \cite{Forrester} taking $\xi=1$ and $t=L k_F/\pi$.
  The dashed line $\textcolor{myblue}{\mydashedlined}$ is the expansion 
  (\ref{tau0_large}) of $\tau_0(Lk_F)$ for large $Lk_F$.}
  \label{fig:efp_xy_gamma_0}
   \end{figure}

Since for $Lk_F\rightarrow 0$, $\mathcal{P}_0\rightarrow 1$ the constant in~\eqref{tau_sine} is fixed to one, we can finally conclude
\begin{equation}\label{tau0_small}
\log\tau_0(Lk_F)\stackrel{Lk_F\ll 1}{\sim} -\frac{Lk_F}{\pi}
-\frac{1}{2}\left(\frac{Lk_F}{\pi}\right)^2
-\frac{1}{3}\left(\frac{Lk_F}{\pi}\right)^3
+\frac{\pi^2-9}{36}\left(\frac{Lk_F}{\pi}\right)^4.
\end{equation}
In~\cite{Gamayun}, it is also determined the behaviour
of $\tau_0$ for large $Lk_F$, 
\begin{equation}\label{tau0_large}
  \log\tau_0(Lk_F)\stackrel{Lk_F\gg 1}{\sim} -\frac{1}{2}\left(\frac{Lk_F}{2}\right)^2
  -\frac{1}{4}\log\frac{Lk_F}{2}+\log\left(\sqrt{\pi}
  G(1/2)^2\right).
\end{equation}
As Dyson already noted in~\cite{Dyson}, the asymptotics
(\ref{efp_omega_0}) predicted by the Widom theorem for
$\mathcal{P}_0(L, k_F)$ leads to (\ref{tau0_large}) when $k_F\ll 1$, 
consistently with the continuum limit $a\to0$.

In fig. \ref{fig:efp_xy_gamma_0}, we check numerically
the double scaling limit (\ref{efp_painleve_omega_0}). We represent $\log\mathcal{P}_0$
as a function of $Lk_F$ for two fixed values of $L$ and varying
$k_F$. The dots correspond to the numerical results obtained
for $\mathcal{P}_0$ diagonalizing the matrix $S$, see
(\ref{matrix_s_omega_0}), and using the eigenvalues to compute
$\mathcal{P}_0$ from (\ref{efp_det}). The solid curve is the
expansion (\ref{tau0_small}) of $\log \tau_0$ for $Lk_F\to0$
while the dashed line represents the asymptotics (\ref{tau0_large})
of $\log\tau_0$ for $L k_F\to \infty$.

\section{Conclusions}\label{sec:conclusions}
\begin{figure}[t]
  \centering
  \includegraphics[width=0.45\textwidth]{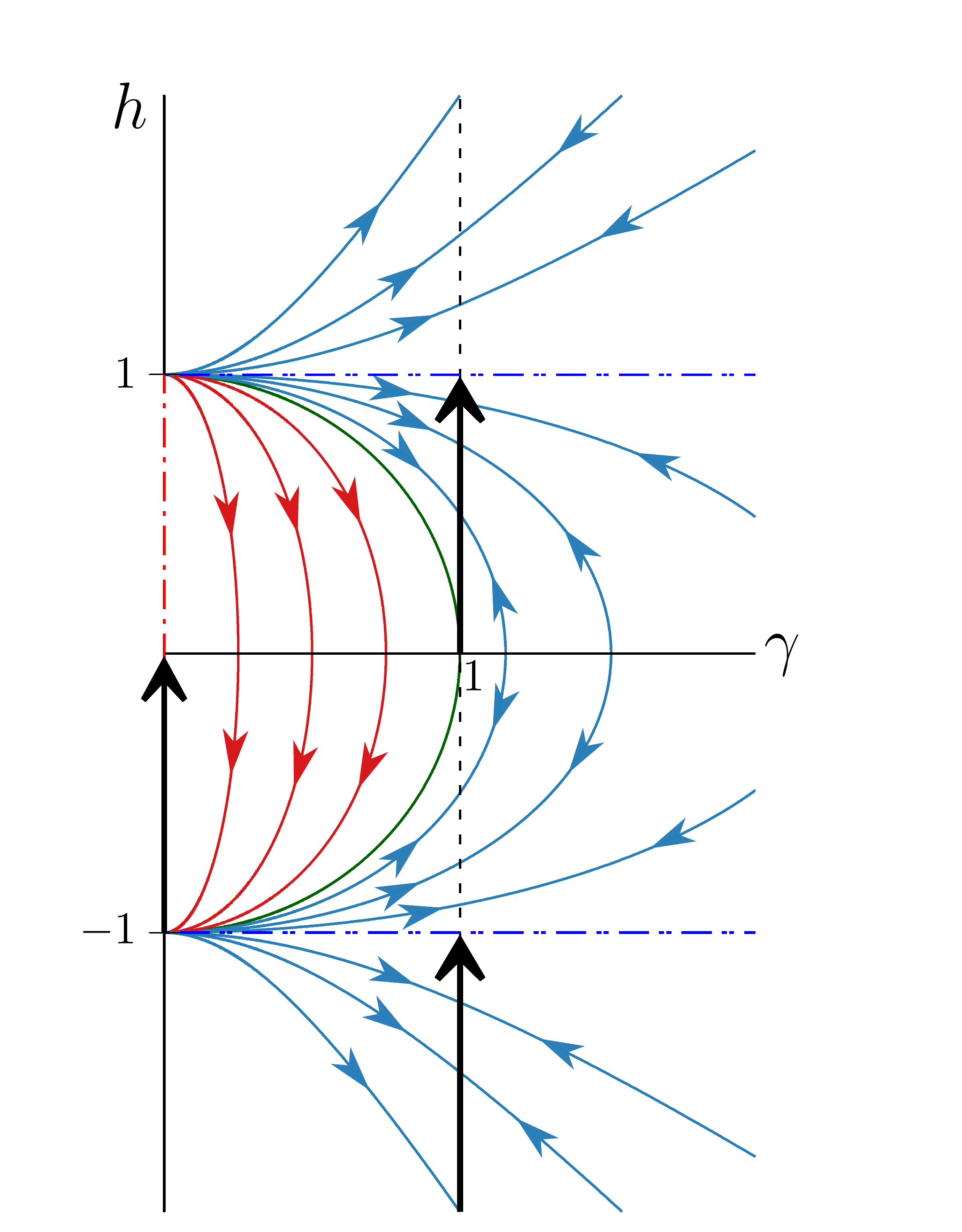}
  \caption{The figure shows the flow in the parameter space of the XY plane associated to the M\"obius transformation~\eqref{moebius} that keeps invariant the ratio $\frac{h^2-1}{\gamma^2}$. The vertical black lines denote schematically the transitions in the double scaling limits described by  the $\tau$ function in sec.~\ref{sec:results} for $\gamma=1$ and in sec.~\ref{sec:xx} for $\gamma=0$.}
  \label{fig:conclusions}
   \end{figure}

In this paper we revisited the problem of determining the emptiness formation probability in the XY spin chain. In particular we complemented the asymptotic results obtained in~\cite{Abanov, Franchini} showing that the emptiness formation probability  in the double scaling limit $L\rightarrow\infty$, $|h|\rightarrow 1$ is the $\tau$ function of a Painlev\'e V equation~\cite{Claeys}. By exploiting the combinatorial representation proposed in~\cite{Gamayun, Lisovyy}  we  determined a power series expansion for the emptiness formation probability around $L\log|h|=0$, which has been tested numerically finding excellent agreement. Our results are exact~\cite{Claeys} at $\gamma=1$, for the quantum Ising chain, and have been extended to $\gamma\not=1$ through a conjecture, which is based~\cite{Ares1} on the symmetry properties of the Toeplitz determinant under a M\"obius transformation in the parameter space of the XY chain.
The M\"obius transformations, which manifest in the large-$L$ limit, can be interpreted  as flows in the parameter space connecting points $(\gamma, h)$, located outside the circle $h^2+\gamma^2=1$, with  points along the Ising line $(\gamma=1, h')$, see fig~\ref{fig:conclusions}. Points inside the circle are attracted instead by the  fixed point $(\gamma=0, -1)$, see again fig.~\ref{fig:conclusions}. The emptiness formation probabilities, as well as the entanglement entropies~\cite{Ares1, Ares2}, are argued to be invariant for theories that sit on the same flow. Although a complete derivation is still lacking, our conjectures have  been tested numerically with remarkable accuracy. 

Moreover, since in the double scaling limit the emptiness formation probability is a function of the scaling variable $L/\xi$, being $\xi$ the correlation length, it seems meaningful to ask whether it could be obtained  directly in a field theory setting. Painlev\'e equations in the massive Ising field theory have  been obtained in the past, see for instance~\cite{Bernard, Bernard2, Doyon}, and applied to the study of correlation functions including the entanglement entropies~\cite{Keating2, Casini}.

Finally, we mention that our results could be generalized to the analysis of the full counting statistics in the XY chain~\cite{Ivanov, Abanov2, Ivanov2}, where the emergence of a similar Painlev\'e V equation is expected when approaching criticality. Analogous expectations apply also to the charged entropies considered recently in~\cite{Moses, Bonsignori}.

\vspace*{1cm}
   %%%%%%%%%%%%%%%%%%%%%%%%%%%%%%%%%%%%%%%%%%%%%%%%%%%%%%%%%%%%%%%%
  %%%%%%%%%%%%%%%%%%%%%%%%%%%%%%%%%%%%%%%%%%%%%%%%%%%%%%%%%%%%%%%
\textbf{Acknowledgments.}
 %%%%%%%%%%%%%%%%%%%%%%%%%%%%%%%%%%%%%%%%%%%%%%%%%%%%%%%%%%%%%%%%
  %%%%%%%%%%%%%%%%%%%%%%%%%%%%%%%%%%%%%%%%%%%%%%%%%%%%%%%%%%%%%%%
We  are grateful to  Alexander Abanov, Dimitry Gangardt and Jean-Marie St\'ephan for discussions  and to Bruno Carneiro da Cunha and F\'abio Novaes for clarifications on the combinatorial expansion. FA and JV acknowledge financial support from the Brazilian Ministries MEC and MCTIC.  

\appendix

\section{Some general results on the asymptotics of Toeplitz determinants}\label{sec:appendixa}
Consider a real, positive symbol $g$ defined on the
unit circle $S^1$ and with entries in $L^1(S^1)$. We shall
denote by $T_L[g]$ the $L\times L$ Toeplitz matrix generated
by $g$. Its entries are given by the Fourier coefficients
of the symbol, i. e. $(T_L[g])_{nm}=g_{n-m}$, with
\begin{equation*}
  g_k=\frac{1}{2\pi}\int_0^{2\pi} g(\theta)\e^{\ii \theta k}{\rm d}\theta.
\end{equation*}

If the symbol $g$ is smooth enough such that
\begin{equation}\label{smoothness_cond}
  \sum_{k=-\infty}^\infty |g_k|+\sum_{k=-\infty}^\infty |k||g_k|^2<\infty,
\end{equation}
then the Szeg\H{o} theorem \cite{Szego} states that the determinant of
$T_L[g]$, that we denote as $D_L[g]$, has the following
asymptotic expansion with the dimension $L$:
\begin{equation}\label{szego}
  \log D_L[g]=(\log g)_0L+E[g]+o(1)
\end{equation}
where
$$E[g]=\sum_{k=1}^\infty k(\log g)_k(\log g)_{-k},$$
and $(\log g)_k$ are the Fourier coefficients of $\log g$.

If the symbol has jump discontinuities and/or zeros, then
the smoothness condition (\ref{smoothness_cond}) is violated and the Szeg\H{o}
theorem (\ref{szego}) does not apply. In this case, the Fisher-Hartwig
conjecture gives the asymptotic behaviour of the determinant.
Suppose that the symbol $g(\theta)$ presents $R$ Fisher-Hartwig singularities
(jump discontinuities and/or zeros) at the points $0\leq \theta_1<
\theta_2<\dots<\theta_R<2\pi$ and it can be uniquely factorized
in the form
\begin{equation}\label{fh_fact}
  g(\theta)=V(\theta)\prod_{r=1}^R(2-2\cos(\theta-\theta_r))^{\alpha_r}
  \e^{\ii\beta_r(\theta-\theta_r-\pi\sign(\theta-\theta_r))},
\end{equation}
with $V(\theta)$ a strictly positive continuous function that satisfies
(\ref{smoothness_cond}), then the Fisher-Hartwig conjecture \cite{Fisher, Basor1} 
(that it is actually a theorem in this case) predicts that
\begin{equation}\label{fh}
  \log D_L[g]=(\log V)_0 L+\sum_{r=1}^R (\alpha^2-\beta^2)\log L
  + E(V, \{\alpha_r\}, \{\beta_r\}, \{\theta_r\})+o(1),
\end{equation}
where
\begin{eqnarray*}
  E(V, \{\alpha_r\}, \{\beta_r\}, \{\theta_r\})
  &=&E[V]+\sum_{r=1}^R\left[\left(-\alpha_r+\beta_r\right)V_+(\theta_r)
    -\left(\alpha_r+\beta_r\right)V_-(\theta_r)\right]\\
  &&-
  \sum_{1\leq r\neq r'\leq R}(\alpha_r+\beta_r)(\alpha_{r'}-\beta_{r'})\log(1-\e^{\ii(\theta_{r}-\theta_{r'})})\\
  &&+\sum_{r=1}^R\log \frac{G(1+\alpha_r+\beta_r)G(1+\alpha_r-\beta_r)}{G(1+2\alpha_r)},
\end{eqnarray*}
and 
$$V_\pm(\theta)=\sum_{k=1}^\infty(\log V)_{\pm k}\ee^{\ii\theta k}.$$

If the symbol admits more than one factorization,
\begin{equation}\label{gfh_fact}
  g(\theta)=V^{(j)}(\theta)\prod_{r=1}^{R} (2-2\cos(\theta-\theta_r))^{\alpha_r^{(j)}}
  \e^{\ii \beta_r^{(j)}(\theta-\theta_r-\pi\sign(\theta-\theta_r))},
\end{equation}
where $j$ is the label of each factorization, 
then Basor and Tracy \cite{Basor2} proposed a generalization of
the Fisher-Hartwig conjecture. According to it,
\begin{equation}\label{gfh}
  D_L[g]\sim \sum_{j\in \mathscr{S}} \e^{(\log V^{(j)})_0 L}L^{\Omega(j)}
  \ee^{E[V^{(j)}, \{\alpha_r^{(j)}\}, \{\beta_r^{(j)}\}, \{\theta_r\}]},
\end{equation}
as $L\to \infty$. Here
$$\Omega(j)=\sum_{r=1}^R (\alpha_r^{(j)})^2-(\beta_r^{(j)})^2,$$
and
$$\mathscr{S}=\{j\, |\, {\rm Re}\, \Omega(j)=\Omega\}, \quad \mathrm{with}
\quad \Omega=\max_{j}{\rm Re}\,\Omega(j).$$

If the symbol has compact support, then the previous results
are not valid and one has to apply the Widom theorem \cite{Widom}. 
Let suppose that the symbol $g$ is supported on a closed interval $\phi\leq 
\theta\leq 2\pi-\phi$ and its restriction to this interval
is a positive function that satisfies (\ref{smoothness_cond}), then   
\begin{equation*}
 \log D_L[g]=\log\left(\cos\frac{\phi}{2}\right)L^2+(\log \tilde{g})_0L-
\frac{1}{4}\log L+\frac{1}{2}E[\tilde{g}]-\frac{1}{4}\log\sin\frac{\phi}{2}
+\log\left(\sqrt{\pi}G(1/2)^2\right)+o(1),
\end{equation*}
where
\begin{equation*}
\tilde{g}(\theta)=g\left(2\arccos\left(\cos\frac{\phi}{2}\cos\theta\right)\right).
\end{equation*}

Another result on Toeplitz determinants that is used 
in the paper is the localization theorem \cite{Basor1}:
If we consider two symbols $g_1$ and $g_2$ such that their 
Toeplitz matrices are invertible for large $L$,
then
\begin{equation*}
\lim_{L\to\infty}\frac{D_L[g_1g_2]}{D_L[g_1]D_L[g_2]}<\infty
\end{equation*}
provided the semi-infinite matrices $T[g_1g_2]-T[g_1]T[g_2]$ and
$T[g_2g_1]-T[g_2]T[g_1]$, obtained from $T_L[\bullet]$ when $L\to\infty$,
are trace-class. These operators are trace-class if there is a smooth partition
of the unit $\{f_1(\theta),f_2(\theta)\}$ such that the derivatives of 
$g_1f_1$ and $g_2f_2$ are H\"older continuous with exponent larger than
1/2. Namely, the operators are trace-class if $g_1$ and $g_2$ have not 
Fisher-Hartwig singularities at the same points.

\section{Remarks on the notation employed}

\label{sec:appendixb}
In~\cite{Gamayun, Lisovyy}, it was deduced the full expansion of the 
$\tau$ function for the Jimbo-Miwa-Okamoto
form of the Painlev\'e V equation 
\begin{equation}\label{pV_gil}
 (x\tilde{\zeta}'')^2=(\tilde{\zeta}-x\tilde{\zeta}'+2(\tilde{\zeta}')^2)^2
 -\frac{1}{4}\left((2\tilde{\zeta}'-\theta_*)^2-4\theta_0^2\right)
 \left((2\tilde{\zeta}'+\theta_*)^2-4\theta_t^2\right).
\end{equation}
In this case, the $\tau$ function is defined as 
\begin{equation}\label{tau_gil}
 \tilde{\zeta}(x)=x\frac{{\rm d}}{{\rm d}x}
 \log\tau(x)-\frac{\theta_*}{2}x-\theta_0^2-\theta_t^2
 -\frac{\theta_*^2}{2}.
\end{equation}

On the other hand, in~\cite{Jimbo},
the first order terms of the expansion of the $\tau$ function
were obtained for a Painlev\'e V equation in the form

\begin{multline}\label{pV_jimbo}
 (x\zeta'')^2=\left(\zeta-x\zeta+2(\zeta')^2-
 (2\theta_0^{\rm J}+\theta_\infty^{\rm J})\zeta'\right)^2
 \\
  -4\zeta'(\zeta'-\theta_0^{\rm J})
 \left(\zeta-\frac{\theta_0^{\rm J}-
 \theta_t^{\rm J}+\theta_\infty^{\rm J}}{2}\right)
 \left(\zeta-\frac{\theta_0^{\rm J}+\theta_t^{\rm J}
 +\theta_\infty^{\rm J}}{2}\right).
\end{multline}
In \cite{Jimbo}, the $\tau$ function for the differential
equation~\eqref{pV_jimbo} is defined as
\begin{equation}\label{tau_jimbo}
 \zeta(x)=x\frac{{\rm d}}{{\rm d}x}
 \log\tau^{\rm J}(x)+\frac{\theta_0^{\rm J}
 +\theta_\infty^{\rm J}}{2}x
 +\frac{(\theta_0^{\rm J}+\theta_\infty^{\rm J})^2-(\theta_t^{\rm J})^2}{4}.
\end{equation}

Observe that the Painlev\'e V equation 
(\ref{painleve_v}) which is considered in 
this paper is of the form (\ref{pV_jimbo}), but 
written in terms of the parameters 
$\theta_0$, $\theta_t$ and $\theta_*$. Furthermore,
the $\tau$ function introduced in section~\ref{sec:expansion} 
is the one in~\cite{Gamayun}. Namely, despite using $\zeta(x)$ rather than $\tilde{\zeta}(x)$, we 
study $\tau(x)$ instead of $\tau^{\rm J}(x)$. The 
equation (\ref{tau_function}) connects $\tau(x)$ with the 
solution of (\ref{pV_jimbo}), $\zeta(x)$. This identity 
can be obtained as follows.

The solutions of eqs. (\ref{pV_gil}) and (\ref{pV_jimbo}) 
can be related by the transformation, see also \cite{BasorBleher},
\begin{equation}\label{jimbo_gil_transf}
 \tilde{\zeta}(x)=\zeta(x)-\left(\theta_0-
 \frac{\theta_*}{2}\right)x
 -2\left(\theta_0-\frac{\theta_*}{2}\right)^2.
\end{equation}
Then it is straightforward to see that 
\begin{equation}\label{jimbo_gil_parameters}
 \theta_0=\frac{\theta_0^{\rm J}}{2},\quad 
 \theta_t=\frac{\theta_t^{\rm J}}{2},\quad
 \theta_*=-\frac{\theta_\infty^{\rm J}}{2},
\end{equation}
or 
\begin{equation*}
 \theta_0=\frac{\theta_t^{\rm J}}{2},\quad 
 \theta_t=\frac{\theta_0^{\rm J}}{2},\quad 
 \theta_*=\frac{\theta_\infty^{\rm J}}{2}.
\end{equation*}
Here we only consider the case (\ref{jimbo_gil_parameters}).
By inserting (\ref{tau_gil}) and (\ref{tau_jimbo}) 
into (\ref{jimbo_gil_transf}), we conclude 
that the $\tau$ functions of eqs. (\ref{pV_gil}) 
and (\ref{pV_jimbo}) are connected by the identity
\begin{equation}\label{tau_jimbo_gil}
\tau^{\rm J}(x)=x^{-\theta_*^2}\tau(x).
\end{equation}
Finally, plugging (\ref{tau_jimbo_gil}) into 
(\ref{tau_jimbo}) we obtain
\begin{equation*}
 \zeta(x)=x\frac{{\rm d}}{{\rm d}x}\log \tau(x)+(\theta_0-\theta_*)x
  +\theta_0^2-\theta_t^2-2\theta_0\theta_*,
\end{equation*}  
which is precisely eq. (\ref{tau_function}).

Note also that, comparing 
the expansion for $\tau(x)$ written in 
(\ref{tau_expansion_jimbo}) with that 
found in Theorem 3.1 of~\cite{Jimbo}, the relation
between the parameter $\sigma$ 
considered here and the one in~\cite{Jimbo}, $\sigma^{\rm J}$, is 
$\sigma^{\rm J}=2\sigma$.


\begin{thebibliography}{XXX}

  \bibitem{Korepin-book} V. Korepin, N. Bogoliubov, A. Izergin, \textit{Quantum Inverse Scattering Methods and Correlation Functions}, Cambridge University Press, (1993)

  \bibitem{Korepin1} V. Korepin, A. Izergin, F. Essler, D. Uglov, \textit{Correlation functions of the spin 1/2 XXX antiferromagnet}, Phys. Lett. A 190 182 (1994), arXiv:cond-mat/9403066
  
  \bibitem{Essler} F. Essler, H. Frahm, A Izergin, V. Korepin, \textit{Determinant Representation for correlation functions of spin 1/2 XXX and XXZ Heisenberg magnets}, Commun. Math. Phys. 174, 191 (1995), arXiv:hep-th/9406133

  \bibitem{Shiroishi} M. Shiroishi, M. Takahashi, Y. Nishiyama,
    \textit{Emptiness formation probability for the one-dimensional isotropic
      XY model}, J. Phys. Soc. Japan 70,  3535 (2001), arXiv: cond-mat/0106062 
  
  \bibitem{Abanov-Korepin} A. G. Abanov, V. Korepin, \textit{On the probability of ferromagnetic strings in antiferromagnetic spin chains}, Nucl. Phys. B 647 565 (2002), arXiv:cond-mat/0206353 
  
  \bibitem{Kitanine1} N. Kitanine, J-M. Maillet, N. Slavnov, V. Terras, \textit{Emptiness formation probability of the spin XXZ spin 1/2 chain at Delta=1/2}, J. Phys. A: Math. Gen. 35 L753 (2002), arXiv:hep-th/0201134
  
  \bibitem{Kitanine2} N. Kitanine, J-M. Maillet N. Slavnov, V. Terras, \textit{Large distance asymptotic behaviour of the emptiness formation probability of the XXZ spin 1/2 Heisenberg chain}, J. Phys. A: Math. Gen. 35 L753 (2003), arXiv:hep-th/0210019
  
  \bibitem{Lukyanov} V. Korepin, S. Lukyanov, Y. Nishiyama, M Shiroishi, \textit{Asymptotic behaviour of the emptiness formation probability in the critcal phase of XXZ spin chain}, Phys. Lett. A 312 21 (2003), arXiv:cond-mat/0210140 [cond-mat.stat-mech]
  
  \bibitem{Kozlowsky} K. Kozlowsky, \textit{On the emptiness formation probability of the open XXZ spin 1/2 chain}, J. Stat. Mech. P02006 (2008), arXiv:0708.0433 [hep-th]
  
  \bibitem{Cantini} L. Cantini, \textit{Finite Size Emptiness formation probability of the XXZ spin chain at Delta=-1/2}, J. Phys. A: Math. Theor. 45 135207 (2012), arXiv:1110.2404 [math-ph]

  \bibitem{Abanov-lectures} A. G. Abanov, \textit{Hydrodynamics of correlated systems, Emptiness Formation Probability and Random Matrices}, arXiv: cond-mat/0504307 (2005)

\bibitem{Stephan} J-M. St\'ephan,
    \textit{Emptiness formation probability, Toeplitz determinants, and conformal field theory},
    J. Stat. Mech. P05010 (2014), arXiv:1303.5499 [cond-mat.stat-mech]  

    
\bibitem{Allegra} N. Allegra, J. Dubail, J-M. St\'ephan, J. Viti, \textit{Inhomogeneous field theory inside the arctic circle}, J. Stat. Mech. P053108 (2016), arXiv:1512.02872 [cond-mat.stat-mech]    

\bibitem{Colomo1} F. Colomo, A. Pronko, \textit{Emptiness formation probability in the domain-wall six-vertex model}, Nucl. Phys. B 798  340-362 (2008), arXiv:0712.1524 [math-ph]

\bibitem{Colomo2} F. Colomo, A. Pronko, \textit{The arctic curve of the domain wall six-vertex model},  J. Stat. Phys. 138 662-700 (2010), arXiv:0907.1264 [math-ph] 

\bibitem{M1} M. A. Rajabpour, \textit{Formation probabilities in quantum critical chains and Casimir effect}, Europhys. Lett. 112 66001 (2015), arXiv:1512.01052 [cond-mat.str-el]  

\bibitem{M2} M. A. Rajabpour, \textit{Finite size corrections to scaling of the formation probabilities and the Casimir effect in the conformal field theories}, J. Stat. Mech 123101 (2016), arXiv:1607.07016 [cond-mat.stat-mech]

\bibitem{M3}  K. Najafi and M. A. Rajabpour, \textit{Formation probabilities and Shannon information and their time evolution after quantum quench in transverse-field XY-chain}, Phys. Rev. B 93, 125139 (2016), arXiv:1511.06401 [cond-mat.str-el]


\bibitem{Cardy} J. Cardy, \textit{Boundary conditions, fusion rules and the Verlinde formula}, Nucl. Phys. B 324 581 (1989)
  
  \bibitem{Abanov} A. G. Abanov, F. Franchini,
   \textit{Emptiness Formation Probability for the Anisotropic XY Spin Chain
    in a Magnetic Field},
   Phys. Lett. A 316 (2003) 342-349, arXiv: cond-mat/0307001

  \bibitem{Franchini} F. Franchini, A. G. Abanov,
    \textit{Asymptotics of Toeplitz determinants and the emptiness formation
      probability for the XY spin chain},
    J. Phys. A: Math. Gen. 38 (2005) 5069-5095, arXiv:cond-mat/0502015
    
    
 \bibitem{LSM} E. Lieb, T. Schultz, D. Mattis, \textit{Two soluble models of an antiferromagnetic chain}, Ann. Phys. 16 407-66 (1961)   

  \bibitem{Szego} G. Szeg\H{o},
    \textit{On certain hermitian forms associated with the Fourier series
     of a positive function}, Festschrift Marcel Riesz, Lund (1952), 228–238

  \bibitem{Fisher} M. E. Fisher, R. E. Hartwig, 
   \textit{Toeplitz determinants, some applications,
      theorems and conjectures}, Adv. Chem. Phys. 15, 333-353 (1968)
      
  \bibitem{Basor1} E. L. Basor,
    \textit{A localization theorem for Toeplitz determinants},
    Indiana Math. J. 28, 975 (1979)

 \bibitem{Basor2} E. L. Basor, C. A. Tracy,
    \textit{The Fisher-Hartwig conjecture and generalizations},
    Physica A 177, 167 (1991)

 \bibitem{Basor3} E. L. Basor, K. E. Morrison,
    \textit{The Fisher-Hartwig conjecture and Toeplitz eigenvalues},
    Lin. Alg. Appl. 202, 129 (1994)
    
 \bibitem{Widom} H. Widom,
    \textit{The Strong Szeg\H{o} limit theorem for circular arcs},
    Ind. Univ. Math. J. 21, 277 (1971)
    
    \bibitem{BPZ} A. Belavin, A. Polyakov, A.B. Zamolodchikov, \textit{Infinite Conformal Symmetry in two-dimensional Quantum Field Theory}, Nucl. Phys. B 241 333 (1984)

\bibitem{Claeys} T. Claeys, A. Its, I. Krasovsky,
  \textit{Emergence of a singularity for Toeplitz determinants and Painlev\'e V},
  Duke Math J. 160 (2011) 207-262, arXiv:1004.3696 [math-ph]

\bibitem{Jimbo} M. Jimbo,
  \textit{Monodromy Problem and the Boundary Condition for Some Painlev\'e
    equations}, Publ. RIMS, Kyoto Univ. 18 (1982), 1137-1161.

 \bibitem{McCoy-book} B. McCoy, T. Wu, \textit{The two-dimensional Ising model}, Harvard University Press: Cambridge MA, 1973

\bibitem{McCoy} T. Wu, B. McCoy, C. Tracy, E. Barouch, \textit{Spin-spin correlation functions for the two dimensional Ising model: exact theory in the scaling limit}, Phys. Rev. B 13 316 (1976)

\bibitem{Tracy-Ising} C. Tracy, \textit{Asymptotics of a tau function arising in the two-dimensional Ising model}, Commun. Math. Phys. 142 297-311 (1991)
    
 \bibitem{Tracy-review} C. Tracy, H. Widom, \textit{Painlev\'e functions in Statistical Physics}, Publ. RIMS Kyoto Univ. 47 361 (2011), arXiv:0912.2362 [math.PR]

    
 \bibitem{Nagoya} H. Nagoya, \textit{Irregular conformal blocks, with an application to the fifth and forth Painlev\'e equations}, J. Math. Phys. 56, 123505 (2015), arXiv:1505.02398 [math-ph]

  \bibitem{Gamayun0} O. Gamayun, N. Iorgov, O. Lisovyy,
  \textit{Conformal field theory of Painlev\'e VI},
  J. High Energ. Phys. (2012) 2012: 38, arXiv:1207.0787 [hep-th]
  
  \bibitem{Gamayun} O. Gamayun, N. Iorgov, O. Lisovyy,
  \textit{How instanton combinatronics solves Painlev\'e VI, V and III's},
  J. Phys A: Math. Theor. 46 (2013) 335203, arXiv:1302.1832 [hep-th] 

  
 \bibitem{Lisovyy} O. Lisovyy, H. Nagoya, J. Roussillon, \textit{Irregular conformal blocks and connection formulae for Painlev\'e V functions},  J. Math. Phys. 59, 091409 (2018), arXiv:1806.08344 [math-ph]
 
 \bibitem{Dyson} F. J. Dyson,
   \textit{Fredholm determinants and inverse scattering problems},
   Comm. Math. Phys. 47, 2, 171 (1976)
   
  \bibitem{Jimbo2} M. Jimbo, T. Miwa, Y. Mori,  M. Sato,
   \textit{Density matrix of an impenetrable Bose gas and the fifth Painlev\'e transcendent},
   Physica D 1, 80 (1980)
   
   \bibitem{Grassi} A. Grassi, J. Gu, 
   \textit{Argyres-Douglas theories, Painlev\'e II and quantum mechanics},
   J. High Energ. Phys. (2019) 2019: 60, arXiv:1803.02320 [hep-th]
   
   \bibitem{BasorBleher} E. Basor, P. Bleher, R. Buckingham, T. Grava, A. Its, E. Its, J. P. Keating
  \textit{A representation of joint moments of CUE characteristic polynomials in terms of Painleve functions},
  arXiv:1811.00064 [math-ph]

  \bibitem{CarneiroCunha}  B. Carneiro da Cunha, J. P. Cavalcante,
  \textit{Confluent conformal blocks and the Teukolsky master equation},
  arXiv:1906.10638 [hep-th]
  

 \bibitem{Vilenkin} N. Ja. Vilenkin, 
 \textit{Special functions and the theory of group representations}, 
 Translations of Mathematical Monographs, 22, AMS (1968)
 
 \bibitem{Ares1} F. Ares, J. G. Esteve, F. Falceto, A. R. de Queiroz,
  \textit{On the M\"obius transformation in the entanglement entropy of fermionic
  chains}, J. Stat. Mech. (2016) 043106, arXiv:1511.02382 [math-ph]
  
 \bibitem{Ares2} F. Ares, J. G. Esteve, F. Falceto, A. R. de Queiroz,
   \textit{Entanglement entropy and Möbius transformations for critical fermionic chains},
   J. Stat. Mech. (2017) 063104, arXiv:1612.07319 [quant-ph]

 
 \bibitem{Krasovsky} I. Krasovsky,
   \textit{Aspects of Toeplitz determinants},
   Progress in Probability 64, 305 (2011), arXiv:1007.1128 [math-ph]

 \bibitem{Ivanov} D. A. Ivanov, A. G. Abanov, V. V. Cheianov,
   \textit{Counting free fermions on a line: a Fisher-Hartwig asymptotic expansion
     for the Toeplitz determinant in the double-scaling limit},
   J. Phys. A: Math. Theor. 46, 085003 (2013), arXiv:1112.2530 [cond-mat.str-el]

 \bibitem{Abanov2} A. G. Abanov, D. A. Ivanov, Y. Qian,
   \textit{Quantum fluctuations of one-dimensional free fermions and Fisher-Hartwig formula for Toeplitz determinants},
   J. Phys. A: Math. Theor. 44, 485001 (2011), arXiv:1108.1355 [cond-mat.str-el]

 \bibitem{Metha_book} M.L. Metha, \textit{Random Matrices}, Academic Press (1991) 
 
 \bibitem{Perk1} B.M. McCoy, J.H.H. Perk, R.E. Shrock,
 \textit{Time-dependent correlation functions of the transverse Ising chain
  at the critical magnetic field}, Nucl. Phys. B 220 (1983) 35-47
  
  \bibitem{Perk2} B.M. McCoy, J.H.H. Perk, R.E. Shrock,
  \textit{Correlation functions of the transverse Ising chain at the critical
  field for large temporal and spatial separations},
  Nucl. Phys. B 220 (1983) 269-282
  
  \bibitem{Perk3} J.H.H. Perk, H. Au-Yang,
  \textit{New Results for the Correlation Functions of the Ising Model
  and the Transverse Ising Chain},
  J. Stat. Phys. 135 (2009) 599-619, arXiv:0901.1931 [cond-mat.stat-mech]
   
   \bibitem{Forrester} P J Forrester,
   \textit{Log-Gases and Random Matrices (London Mathematical Society Monographs)},
     Princenton University Press (2010)

\bibitem{Bernard} O. Babelon, D. Bernard, \textit{From Form Factors to correlation functions: The Ising model}, Phys. Lett. B 288: 113-120 (1992), arXiv:hep-th/9206003

\bibitem{Bernard2} D. Bernard, A. LeClair, \textit{Differential equations for sine-Gordon correlation funtions at the free fermion point}, Nucl. Phys. B 426 534-58 (1994), arXiv:hep-the9402144 
     
\bibitem{Doyon} B. Doyon, \textit{Two-point correlation functions of scaling fields in the Dirac theory on the Poincar\'e disk}, Nucl. Phys. B 675 607-30 (2003), arXiv:hep-th/0304190
  
\bibitem{Keating2} J. Keating, F. Mezzadri, \textit{Random Matrix Theory and Entanglement in Quantum Spin Chains}, Commun. Math. Phys. Vol. 252, 543-579 (2004), arXiv:quant-ph/0407047    
    
\bibitem{Casini} H. Casini, C. D. Fosco, M. Huerta, \textit{Entanglement and $\alpha$ entropies for a massive Dirac field in two dimensions}, J. Stat. Mech. 0507:P07007 (2005), arXiv:cond-mat/0505563

\bibitem{Ivanov2} D. A. Ivanov, A. G. Abanov, \textit{Characterizing correlations with full counting statistics: classical Ising and quantum XY spin chains}, Phys. Rev. E 87, 022114 (2013), arXiv:1203.6325 [cond-mat.str-el]

\bibitem{Moses} M. Goldstein, E. Sela, \textit{Symmetry-Resolved Entanglement in Many-Body Systems}, Phys. Rev. Lett. 120, 200602 (2018), arXiv:1711.09418 [cond-mat.stat-mech]

\bibitem{Bonsignori} R. Bonsignori, P. Ruggiero, P. Calabrese, \textit{Symmetry resolved entanglement in free fermionic systems}, arXiv:1907.02084 [cond-mat.stat-mech]


     
\end{thebibliography}
\end{document}